\documentclass[useAMS,usenatbib,usegraphicx]{mn2e}
\usepackage{graphicx}

\title[$C_2$ interstellar molecule]
{Abundances and rotational temperatures of  $C_2$
interstellar molecule toward six reddened early type stars
\thanks{Based on observations made with ESO Telescope at the Paranal
Observatory under programme ID 266.D-5655(A), 67.C-0281(A), 71.C-0513(C), and
67.D-0439(A).
}}
\author[M. Ka{\'z}mierczak et al.]
{M. Ka{\'z}mierczak,$^{1}$
\thanks{e-mail: kazmierczak@astri.uni.torun.pl}
M. R. Schmidt,$^{2}$
\thanks{e-mail: schmidt@ncac.torun.pl}
A. Bondar$^{3}$
\thanks{email: arctur@inet.ua}
J. Kre{\l}owski$^{1}$
\thanks{e-mail: jacek@astri.uni.torun.pl}
\\
\medskip \\
$^1$ Centre for Astronomy, Nicolaus Copernicus University,
Gagarina 11, 87-100 Toru{\'n}, Poland\\
$^2$ Nicolaus Copernicus Astronomical Center, 
ul. Rabia{\'n}ska 8, 87-100 Toru{\'n}, Poland\\
$^3$ International Centre for Astronomical 
and Medico-Ecological Research, Terskol, Russia}

\date{Accepted...
      Received ...
      in original form ...}

\pagerange{\pageref{firstpage}--\pageref{lastpage}}
\pubyear{2008}

\begin{document}

\maketitle

\label{firstpage}

\begin{abstract}

Using high resolution ($\sim$85,000) and high signal-to-noise ratio
($\sim$200) optical spectra acquired with ESO/UVES spectrograph, we
determined interstellar column densities of $C_2$ for six Galactic lines of
sight with E(B-V)'s ranging from 0.33 to 1.03. For the purpose we identified
and measured absorption lines belonging to the (1,0), (2,0) and (3,0)
Phillips bands $\rm A^{1}\Pi_u-X^{1}\Sigma^{+}_g$. Identification of a few
lines of the $C_2$ (4,0) Phillips system towards HD\,147889 is reported. The
curve of growth method is applied to equivalent widths for the determination
of column densities of individual rotational levels of $C_2$. Excitation
temperature is extracted from the rotational diagrams. Physical parameters
of the intervening molecular clouds: gas kinetic temperatures and densities
of collision partners were estimated from comparison with the theoretical
model of excitation of $C_2$ \citep{Dishoeck1982}.

\end{abstract}

\maketitle

\section{Introduction}
Diatomic carbon is found in a variety of celestial bodies such as comets
($C_2$ Swan band - \cite{Mayer1968}; \cite{Lambert1983}; Ballik-Ramsay and
Phillips bands - \cite{Johnson1983}; \cite{Gredel1989}), the Sun (Swan band
- \cite{Grevesse1973}; \cite{Lambert1978}), in carbon stellar atmospheres
(Swan, Ballik-Ramsay and Phillips bands - \cite{Querci1971}), circumstellar
shells of carbon rich post-AGB stars (Phillips and Swan bands -
\cite{Bakker1996}; \cite{Bakker1997}) and in interstellar clouds.

In the interstellar medium, $C_2$ was first detected by \cite{Souza1977}.
They observed absorption lines of the \mbox{(1,0)} Phillips band $\rm
A^{1}\Pi_u-X^{1}\Sigma^{+}_g$. Lines of the (2,0) Phillips system were first
tentatively detected in interstellar clouds by \cite{Chaffee1978}. The
evident demonstration of the band was shown by \cite{Chaffee1980}. The
Phillips band (3,0) was discovered only by \cite{Dishoeck1986}. Some $C_2$ 
lines  of the (4,0) Phillips band in direction to HD\,204827 were shown by 
\cite{Hobbs2008} (few of very weak and difficult to measure lines of this 
band are demonstrated here for the first time toward HD\,147889).

$C_2$ is particularly interesting because it is the simplest multi-carbon
molecule. Its abundances give information on the chemistry of interstellar
clouds, especially on the pathway to the formation of long chain carbon
molecules which may be connected with carriers of diffuse interstellar bands
(DIBs) \citep{Douglas1977,Thorburn2003}. Additionally, 
the analysis of $C_2$ lines allows to
determine physical conditions in interstellar clouds.

In the interstellar medium, because of the very rare collisions and low
temperatures, we expect only absorption lines from the ground electronic
state. For $C_2$, there are Mulliken (discovered by \cite{Snow1978}, $D ^1
\Sigma^{+}_{u}-X^{1}\Sigma^{+}_g$, $\sim$2313\,\AA, see also -
\cite{Lambert1995,Sonnentrucker2007}), F-X (first detected by
\cite{Lien1984}, $\rm F^{1}\Pi_u-X^{1}\Sigma^{+}_g$, $\sim$1342\,\AA) and
Phillips systems ($A ^1 \Pi_{u}-X ^1 \Sigma^{+}_{g}$,
$\sim$~6900-12000\,\AA).

$C_2$, as a homonuclear diatomic molecule, has a negligible dipole moment
and hence radiative cooling of the excited rotational levels may go only
through the slow quadrupole transitions \citep{Dishoeck1982}. The rotational
levels are pumped by the galactic interstellar radiation field and excited
effectively above the gas kinetic temperature. The rotational ladder of the
electronic absorptions from the high rotational levels (here up to J''=26)
are usually observed. Because of that, lines of the diatomic carbon from a
long-lived ground state rotational levels are measurable and can be the
sensitive diagnostic probes of conditions in molecular clouds that produce
the interstellar absorption lines, in contrast to polar molecules, such as
$CH$ or $CN$, where usually only a few absorption lines from the lowest
rotational levels are observed.

Relative abundances of $C_2$ were predicted for the interstellar clouds on
the basis of detailed chemical models \citep{Black1977,Black1978}. The
excitation mechanisms of $C_2$ have been already analysed in detail
\citep{Chaffee1980, Dishoeck1982}.

The main purpose of this paper is to determine basic physical parameters
like temperatures and densities of the intervening clouds. Four out of the
six analysed objects (Table 1) were studied before, but high quality spectra
from the ESO archive \citep{Bagnulo2003} allowed to accurately measure these
weak features. Broad spectral range of the UVES spectra allows us to analyse
up to four bands by contrast with previous analyses usually
based on single (2,0) band of the Phillips system. Simultaneous observations
of different bands give possibility to compare individually determined
results for each band and make results more reliable. In this paper two new
objects with the interstellar absorption lines of $C_2$ are presented,
increasing the presently available
sample of interstellar clouds (24 lines of sight - see
\cite{Sonnentrucker2007} - Table 13) where a detailed analysis of
excitation of $C_2$ was made (estimates of densities and excitation
temperatures).

The next section describes our data, the observations, reduction and
criteria for choosing stars. In Sect. 3 we introduce methods of analysis of
the observational data. General discussion and summary of our conclusions
are given in Sect. 4 and 5. Discussion of individual results for each star
of the sample and comparison between our results and those of previous
papers is demonstrated in the Appendix.

\section{The observational data}

We used the archived spectroscopic observations collected using the
spectrograph UVES \citep{uves}. UV-Visual Echelle Spectrograph (UVES) is the
high-resolution spectrograph of the VLT fed by the Kueyen telescope of the
ESO Paranal Observatory, Chile. UVES is a cross-dispersed echelle
spectrograph designed to operate with high efficiency from the atmospheric
cut-off at 3000\,\AA~to the long wavelength limit of the CCD detectors (about
11,000\,\AA) so it is really suitable instrument allowing to observe four
bands of the Phillips system in single exposure. The high-quality of UVES
allows to get excellent spectra with the maximal resolution of about 85,000
(accurate value in Table 1) and signal-to-noise ratio above 200.

\begin{table}
\centering
\begin{minipage}{70mm}
\caption{Basic data for program stars (Thorburn et al. 2003, Hunter et al. 2006).}
\label{Table1.}
\begin{tabular}{@{}lcrccc}
\hline
object & name& Sp/L & V & E(B-V) & R \\
\hline
 HD\,76341  &          & O9Ib &  7.17  & 0.49 &79,000 \\
 HD\,147889 &          & B2V  &  7.90  & 1.07 &98,000 \\ 
 HD\,148184 &$\chi$ Oph& B2V  &  4.28  & 0.52 &81,000 \\
 HD\,163800 &          & O4V  &  7.01  & 0.60 &82,000 \\
 HD\,169454 & V430 Sct & B1Ia &  6.65  & 0.93 &110,000\\
 HD\,179406 & 20 Aql   & B3V  &  5.34  & 0.33 &63,000\\
\hline
\end{tabular}
\end{minipage}
\end{table}

We have chosen spectra with interstellar molecular lines of $C_2$ towards
six early-type stars (Table 1) from the UVES archive. Objects with
intermediate colour excess and with one really dominant velocity
component, at the resolution of observational material, 
were selected to make it likely that $C_2$ molecular lines originate from
single clouds. In addition KI 7699\,\AA~line profiles were checked for
possible existence of more than one dominating Doppler components.
Generally we cannot exclude existence of multiple closely-spaced components.
For example, \cite{Welty2001} show a very weak Doppler component in KI and CH toward
HD\,148184 which are not visible in our spectra. 
Presence of weak components should not contaminate rather weak
$C_2$ lines. These authors also show that almost all of the program objects
have multiple components in NaI; however this component may be not related
to molecular ones \citep{Bondar2007}. In ultra-high resolution spectra of
HD\,169454, \cite{Crawford1997} and \cite{Crawford1996} see two Doppler components in
$C_2$ lines. These are yet unresolved at the UVES resolution.

The spectral analysis was made with the Dech20T code \citep{Galazutdinov1992}.

Spectral regions with interstellar $C_2$ absorption lines, especially the
(3,0) and (4,0) bands, are contaminated by atmospheric lines. Our spectra
were divided by spectrum of proper standard (Spica) to remove telluric
lines. In this way a majority of telluric features was removed. Remaining
lines were cut out one by one manually. The continuum was traced to remove
broad stellar absorption features. Also possible broad DIB at 7721.7\,\AA\ 
\citep{Herbig91} blended with $C_2$ (3,0) Q(2) line was removed in this procedure.

\begin{figure*}
\centering
\includegraphics[width=17cm,height=5.7cm]{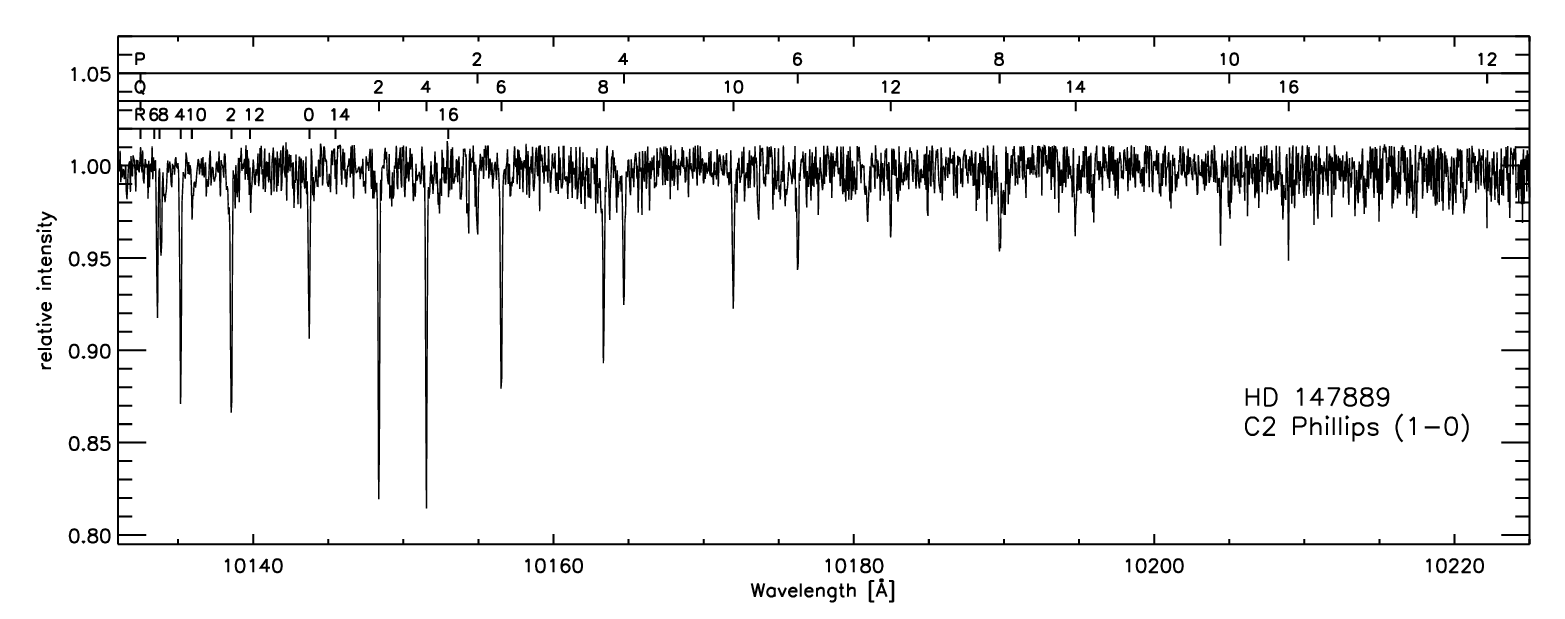}\\
\includegraphics[width=17cm,height=5.7cm]{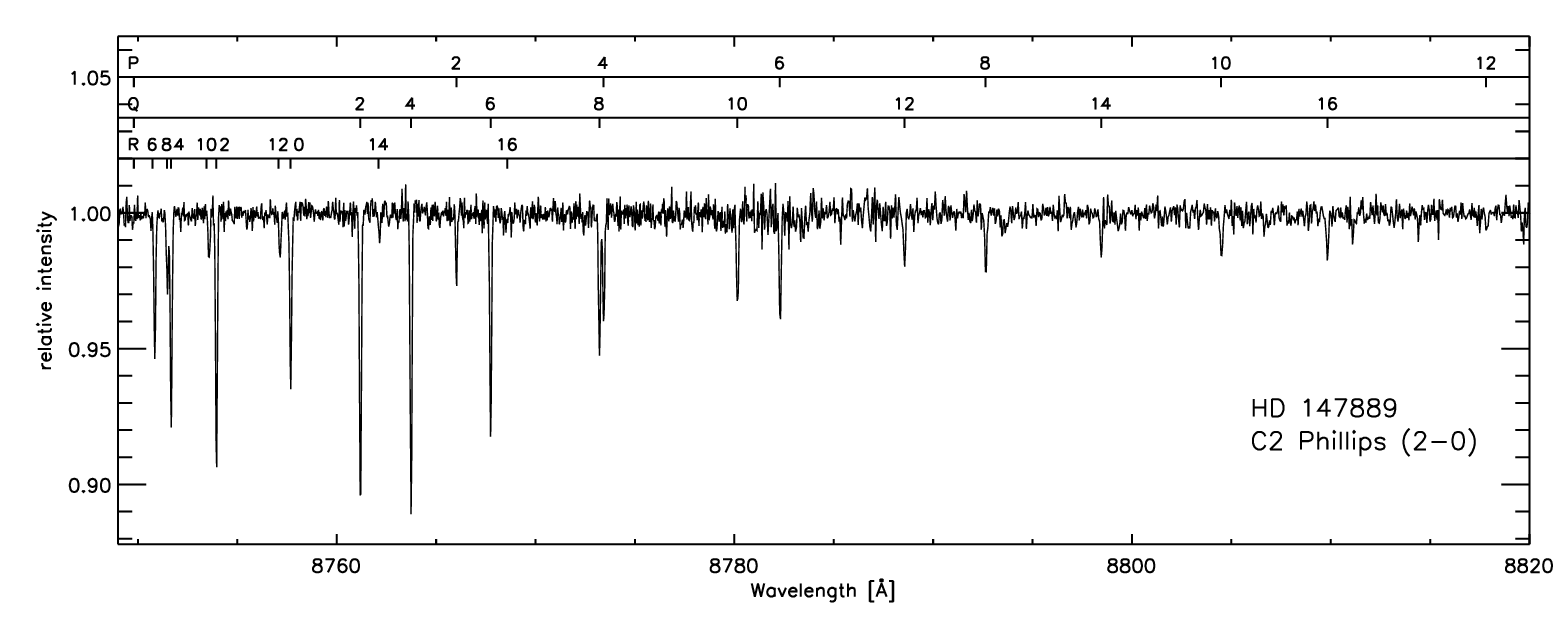}\\
\includegraphics[width=17cm,height=5.7cm]{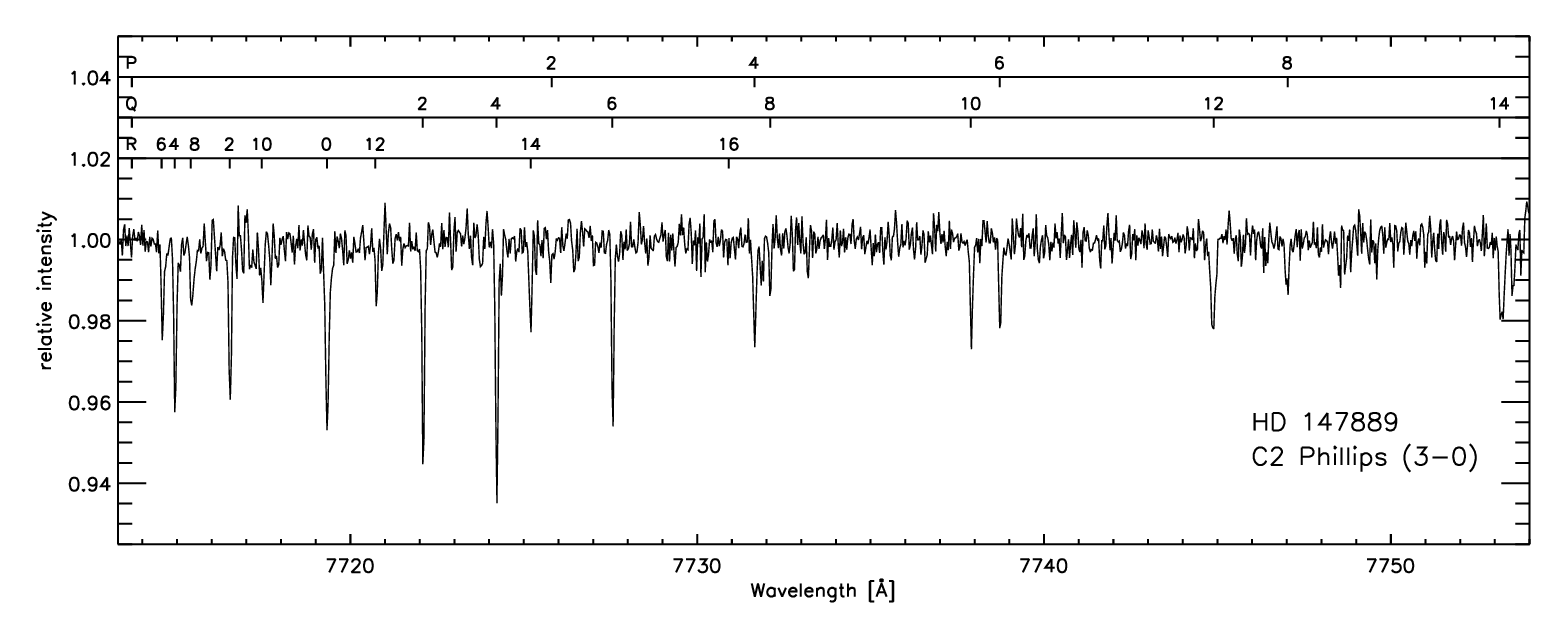}\\
\includegraphics[width=17cm,height=5.7cm]{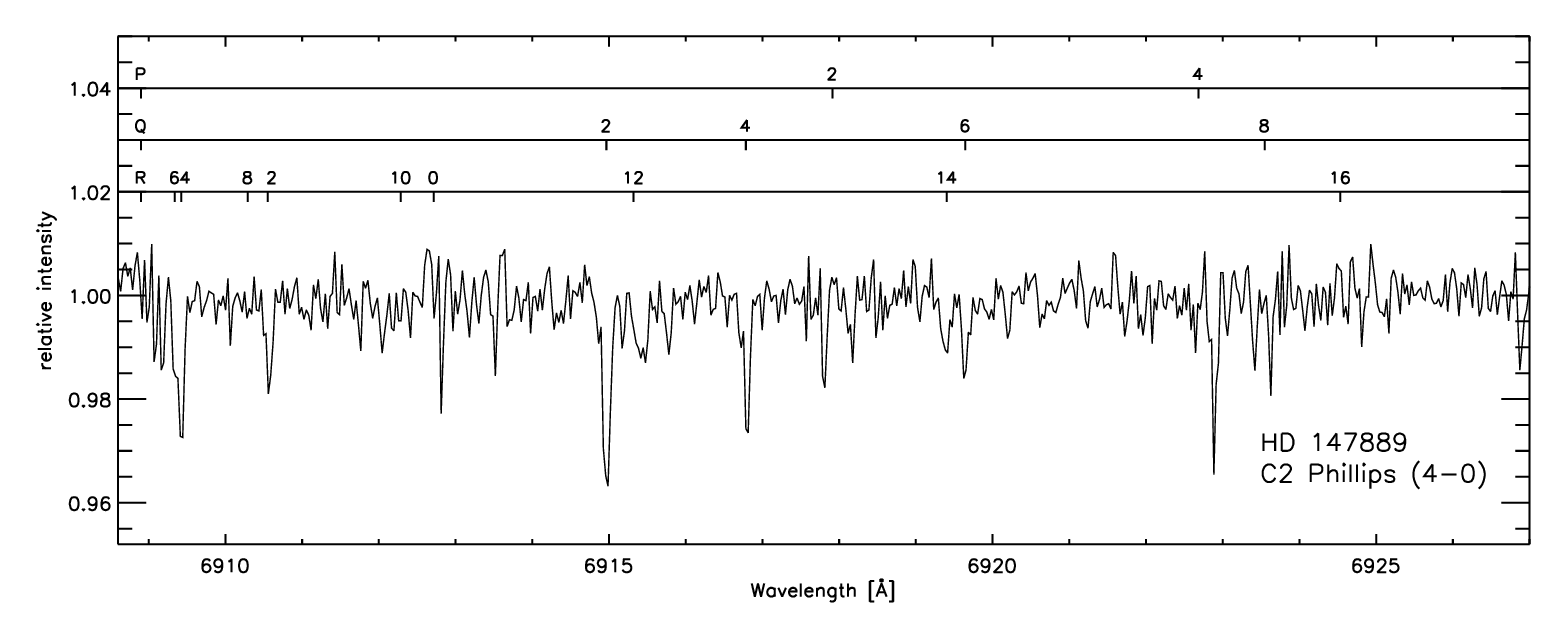}
\caption{The regions of the $C_2$ (1,0), (2,0), (3,0) and (4,0) Phillips band in the spectrum of HD 147889. Spectrum was normalised to a continuum level of 1. $C_2$ absorption lines are indicated. The wavelength scale has been shifted to the rest wavelength velocity frame using the interstellar KI line \mbox{(7698.965 \AA).}}
\label{fig1}
\end{figure*}

The final spectra were normalised to unity to enable measurements of the
equivalent widths. The equivalent widths were measured by fitting Gaussian
profile to each absorption line. At the resolution of the spectra ($\sim
3 - 5\,\rmn{km}\,\rmn{s}^{-1}$) the profiles of single molecular lines would have the shape
of the instrumental profile. Expected Doppler broadening of the profiles
caused by thermal and turbulent motions ($< 0.7\,\rmn{km}\,\rmn{s}^{-1}$ - see
\cite{Crawford1997} (Table 4.)) are significantly lower.

The uncertainties of the equivalent widths ($\Delta EW$)
were estimated with the {\it IRAF}\footnote{
The Image Reduction and Analysis Facility ({\em IRAF}) is distributed by the
National Optical Astronomy Observatories, which is operated by the
Association of Universities for Research in Astronomy, Inc. (AURA), under
cooperative agreement with the National Science Foundation.} 
splot task taking into consideration 
signal-to-noise ratio in the portion of spectrum close to measured 
$C_2$ line. In these cases, where continuum tracing was uncertain,
additional uncertainty was estimated by changing level of continuum.
The errors were propagated to the uncertainties of the
determined parameters (e.g. excitation temperatures, column densities) 
and used in search of the best fit model parameters.

In summary, we measured absorption lines (P, Q, R branches) in bands (1,0)
\mbox{10133 - 10262 {\AA}}, (2,0) \mbox{8750 - 8849 {\AA}}, (3,0) \mbox{7714
- 7793 {\AA}}. In one case, HD\,147889, we were able to identify and measure
interstellar absorption lines of the (4,0) \mbox{6909 - 6974 {\AA}} band
\mbox{(Fig. 1).}

The final spectra of HD\,147889, with the continuum level normalised to
unity, are shown in Figure 1. Broad stellar lines were removed in the
procedure of continuum tracing. There are plenty of weak absorption lines
(Fig. 1) thanks to the high quality of the UVES data.
Even the P(4) and Q(8) lines, which usually are blended in the (2,0) band,
are well separated in all objects.
A lot of $C_2$ lines were also found in the (3,0) band. The oscillator strengths
for (4,0) are about five times weaker than those in the (2,0) and, in
addition, molecular lines around 6915\,\AA\, are seriously contaminated by
telluric lines. $C_2$ lines of the Phillips system, seen in all spectra of
the program stars, allow to identify rotational components of P, Q and R
branches to high rotational levels, up to J''=16. In the spectrum of HD\,147889
we managed to identify Q components up to J"=26.

\section{Results and interpretation}

We have identified lines of the (1,0), (2,0), (3,0) and in one case (4,0)
bands of the $C_2$ Phillips system \mbox{($\rm
A^{1}\Pi_u-X^{1}\Sigma^{+}_g$)}. The equivalent widths with errors of all
measured interstellar lines of $C_2$ towards six program stars are given in
Tables 2-5. Poor quality measurements of equivalent widths are marked with
sign {\em u} behind the value. These uncertain values were not used in the
further analysis of the column densities.

Column density of a rotational level J'' may be derived from the equivalent
width $W_{\lambda}$[m\AA] of the single absorption line using the
relationship \citep{Frisch1972}
\begin{equation}
N_{col} = 1.13 \times 10^{17} {\frac {W_{\lambda}}{f_{ij} \lambda^2}} ~,
\end{equation}
where $\lambda$ is the wavelength in [\AA], $f_{ij}$ is the absorption
oscillator strength. 

The energies of the lower rotational level were determined using
molecular constants of \cite{Marenin1970}. The wavelengths
are generally determined from laboratory wavenumbers 
of \cite{Chauville1977} and \cite{BallikRamsay1963}
converted to air wavelengths using Edlen's formula following \cite{Morton1991}.
Wavelengths of three lines R(2), P(2) and P(4) of the (2,0) band,
absent in \citep{Chauville1977}, were computed with 
\cite{Douay1988} spectroscopic constants. According to
\cite{Douay1988} the line positions calculated with their constants should
be more accurate than the previous measurements.
The oscillator strengths correspond to vibrational oscillator strengths
$f_{10} = 2.38 \times 10^{-3} $, $f_{20} = 1.44 \times 10^{-3}$, 
$f_{30} = 6.67 \times 10^{-4}$, $f_{40} = 2.71 \times 10^{-4}$.
The oscillator strengths for individual transitions were computed 
according to description in \cite{Bakker1996} using their code MOLLEY.
Vibrational oscillator strengths were taken from \cite{Langhoff1990}
for (1,0) and (2,0), \cite{Bakker1997} (citing Langhoff) (3,0) and from
\cite{Dishoeck1983} for (4,0). Source of band origins were 
\cite{Chauville1977} and \cite{BallikRamsay1963b}.

Equation (1) is accurate for the optically thin case (when the absorption
lines are on the linear part of the curve of growth). Only a few lines
from the (1,0) band of HD\,147889, where the equivalent widths are the
largest, are evidently optically thick. Then curve of growth
method was applied for the derivation of column densities instead of the
equation (1). For an optically thick line we determined the turbulent velocity
through the minimalization of the dispersion of column densities for each level.
We checked various values of the velocity dispersion parameter
\mbox{($b=\,0;\,0.5;\,1;\,1.5\,\rmn{km}\,\rmn{s}^{-1}$)} and 
\mbox{0.5\,$\rmn{km}\,\rmn{s}^{-1}$} was
found to give the lowest dispersion. This value was then
applied to all of the program stars. It is consistent with the value derived
in other studies of molecular absorptions (e.g. \cite{Gredel1991}; 
\cite{Crawford1997}). The application of equation (1) to optically thick lines
underestimates column densities by 28 per cent
in the worst case of Q(2) (1,0) absorption line in HD~147889.

The resulting column densities for each rotational level $N_{col}(J'')$,
derived uncertainties and the number of measurements used to determine that value
are presented in Table 6. 
The total $C_2$ column density, defined as the sum of the mean column densities
of the observed levels and of the contribution of the unobserved
levels estimated from the theoretical model characterised by the best-fit parameters
(see below), are shown in column 5 of Table 7.
There were available three
Phillips bands so the column density could be maximally estimated from 9
measurements. Not every part of spectrum allows to measure very weak
interstellar features of $C_2$. In some cases column density was derived
from only one transition (e.g. for the highest rotational levels J''=10, 12,
14, 16). Absorption lines from lower levels (J''= 2 and 4) are the most
populated and the easiest to be measured accurately.

Following \cite{Dishoeck1982} we present derived column densities in
form of rotational diagrams (\mbox{Fig. 2})
where weighted relative column densities \mbox{$-ln[5N_{col}(J'')/(2J''+1)N_{col}(2)]$} 
are plotted versus energy of lower level E''/k (where E'' is the energy 
of the rotational level J'' and k is the Boltzmann constant). The errorbars
correspond to the derived uncertainties. The slope of a straight line 
on this diagram is nicely connected to the excitation temperature, $a=-1 / T_{exc}$.
It is well known from previous works (e.g. \cite{Dishoeck1982})
that populations of all rotational levels cannot be characterised by a
single rotational temperature. The lowest J'' levels are described by the
lower excitation temperature than higher levels. Such behaviour of the
rotational levels was nicely described in the model of excitation of $C_2$
by \cite{Dishoeck1982}. In this model, the molecule is heated by the
electronic transitions from the ground state, and subsequently cooled down
through cascading to the ground electronic state through excited vibrational
levels. Because $C_2$ is a homonuclear species, the mechanism of cooling is
inefficient and the high rotational levels are significantly populated. The
population of the lowest rotational levels is influenced by the collisions
with the gas, mainly atomic and molecular hydrogen (hence the density of
collision partners $n_c = n_H + n_{H_2}$).

\begin{table*}
\centering
\caption{Observation summary table with equivalent widths [m\AA] of $C_2$
(1,0) Phillips lines toward program stars. In table are also $B(N''=J'')$ -
branch identification ($J''$ - low rotational level) and $\lambda$ -
wavelength in air in \AA~(see text for the references).
}
\label{Table 2.}
\begin{tabular}{@{}cccccccc}
\hline
$B(N''=J'')$ &  HD76341 & HD147889   & HD148184    & HD163800 & HD169454 & HD179406&$\lambda$ [{\AA}]\\
\hline  
 $R(6)$ &     	      & $ 8.3\pm1.5$ & $1.6\pm1.0$ & $0.6\pm0.5u$ &  $4.0\pm1.0$ & $3.2\pm1.0 $ & 10133.603 \\
 $R(8)$ &             & $ 5.8\pm1.5$ & $2.4\pm1.5$ & $1.0\pm0.7$  &  $0.7\pm1.3$ & $1.9\pm0.8u$ & 10133.854 \\ 
 $R(4)$ & $2.2\pm1.1$ & $15.6\pm1.4$ & $6.5\pm1.6$ & $3.6\pm1.3$  &  $7.0\pm1.2$ & $8.7\pm1.1 $ & 10135.149 \\
 $R(10)$&     	      & $ 3.4\pm2.2$ &             & $0.5\pm0.5u$ &              & $2.0\pm1.4u$ & 10135.923 \\
 $R(2)$ & $3.2\pm1.2$ & $17.1\pm1.4$ &             & $2.6\pm1.0$  & $12.5\pm1.1$ & $6.1\pm1.5 $ & 10138.540 \\
 $R(12)$&     	      &              & $0.5\pm0.9$ &              &              &              & 10139.805 \\
 $R(0)$ & $1.6\pm1.0$ & $12.4\pm1.4$ & $4.0\pm1.2$ & $3.8\pm1.0$  & $14.3\pm1.0$ & $5.2\pm1.0$  & 10143.723 \\
 $R(14)$&     	      &              &             & $0.8\pm0.5$  &              &              & 10145.505 \\
 $Q(2)$ &     	      & $22.1\pm1.4$ & $6.4\pm0.5$ & $4.3\pm1.3u$ & $21.0\pm1.4$ & $9.0\pm1.3$  & 10148.351 \\
 $Q(4)$ & $2.6\pm1.2$ & $20.3\pm1.6$ & $7.2\pm1.2$ & $3.2\pm1.4u$ & $12.5\pm1.0$ &$11.4\pm1.2$  & 10151.523 \\
 $P(2)$ &             & $ 5.6\pm1.4$ &             & $2.0\pm0.9$  & $ 3.9\pm1.2$ &              & 10154.897 \\
 $Q(6)$ &             & $17.1\pm1.4$ & $4.4\pm1.3$ & $6.0\pm1.4u$ & $ 5.9\pm1.0$ & $6.3\pm1.0$  & 10156.515 \\
 $Q(8)$ & $0.7\pm0.7$ & $13.7\pm1.2$ & $2.8\pm1.5$ & $4.6\pm1.3$  & $ 4.5\pm1.2$ & $4.2\pm1.2$  & 10163.323 \\
 $P(4)$ & $0.3\pm0.3u$& $ 8.8\pm1.5$ & $2.1\pm0.9$ &              & $ 2.8\pm1.0$ &              & 10164.763 \\
 $Q(10)$& $0.8\pm0.8$ & $10.6\pm1.6$ & $2.6\pm1.2$ &              & $ 1.7\pm0.8$ & $3.0\pm1.0$  & 10171.963 \\
 $P(6)$ &     	      & $ 8.8\pm1.5$ & $1.2\pm0.9$ &              & $ 3.0\pm0.9$ &              & 10176.252 \\
 $Q(12)$&     	      & $ 4.7\pm1.2$ &             &              & $ 0.9\pm0.9u$&              & 10182.434 \\
 $P(8)$ &             & $ 4.9\pm1.3$ & $1.4\pm0.7$ &              & $ 1.3\pm0.6$ &              & 10189.693 \\
 $Q(14)$&      	      & $ 3.3\pm1.2$ &             &              &              &              & 10194.755 \\
 $P(10)$&     	      & $ 2.8\pm1.6$ &             &              &              &              & 10204.998 \\
 $Q(16)$&     	      & $ 4.0\pm1.6$ &             &              &              &              & 10208.931 \\
 $P(12)$&     	      & $ 1.8\pm0.9$ &             &              &              &              & 10222.171 \\
 $P(14)$&     	      & $ 1.7\pm0.8$ &             &              &              &              & 10241.246 \\
 $Q(18)$&    	      & $ 1.8\pm1.4u$&             &              &              &              & 10224.982 \\  
\hline
\end{tabular}
\end{table*}
\begin{table*}
\centering
\caption{The same as in Table 2. but for the (2,0) Phillips band.
Wavelengths marked with 1 were computed from spectroscopical constants of Douay et al. 1988.}
\label{Table 3.}
\begin{tabular}{@{}cccccccc}
\hline
$B(N''=J'')$& HD76341 & HD147889     & HD148184    & HD163800     & HD169454 & HD179406 &$\lambda$[{\AA}]\\
\hline  
 $R(6)$ & $0.6\pm0.4$ & $ 6.1\pm0.3$ & $0.7\pm0.4$ & $1.6\pm0.5$  & $ 2.4\pm0.3$ & $1.0\pm0.5$  & 8750.847 \\
 $R(8)$ &             & $ 3.6\pm0.3$ & $0.6\pm0.5$ & $1.0\pm0.4$  & $ 0.7\pm0.3$ & $0.4\pm0.4$  & 8751.487 \\ 
 $R(4)$ & $0.9\pm0.5$ & $ 9.2\pm0.3$ & $1.5\pm0.4$ & $2.0\pm0.5$  & $ 3.7\pm0.3$ & $2.7\pm0.5$  & 8751.684 \\
 $R(10)$&             & $ 2.2\pm0.3$ & $0.7\pm0.4$ &              & $ 0.4\pm0.3u$&              & 8753.578 \\
 $R(2)$ & $1.4\pm0.5$ & $10.5\pm0.3$ & $2.7\pm0.4$ & $1.8\pm0.4$  & $ 8.1\pm0.3$ & $4.1\pm0.5$  & 8753.945$^1$ \\
 $R(12)$&             & $ 1.8\pm0.3$ &             &              & $ 0.3\pm0.2$ &              & 8757.127 \\
 $R(0)$ & $1.1\pm0.4$ & $ 7.5\pm0.3$ & $1.4\pm0.4u$& $1.8\pm0.4$  & $ 8.5\pm0.3$ & $3.3\pm0.4$  & 8757.683 \\
 $Q(2)$ & $1.2\pm0.5$ & $12.7\pm0.3$ & $3.0\pm0.4$ & $2.7\pm0.5$  & $10.3\pm0.3$ & $4.7\pm0.4$  & 8761.194 \\
 $R(14)$&             &              &             &              &              & $0.9\pm0.3u$ & 8762.144 \\
 $Q(4)$ &             & $13.1\pm0.3$ & $3.3\pm0.5$ & $3.5\pm0.7$  & $ 7.9\pm0.3$ & $7.1\pm0.5u$ & 8763.751 \\
 $P(2)$ &             & $ 2.4\pm0.3$ & $0.5\pm0.5u$& $0.5\pm0.4$  & $ 2.0\pm0.3$ & $1.4\pm0.4$  & 8766.026$^1$ \\
 $Q(6)$ &             & $ 9.2\pm0.3$ & $3.4\pm0.4$ & $2.1\pm0.5$  & $ 3.0\pm0.3$ & $3.6\pm0.4$  & 8767.759 \\
 $R(16)$&             &              &             &              &              &              & 8768.628 \\
 $Q(8)$ & $0.3\pm0.3$ & $ 6.3\pm0.3$ & $1.7\pm0.4$ & $1.2\pm0.5$  & $ 2.3\pm0.3$ & $1.3\pm0.8u$ & 8773.220 \\
 $P(4)$ &             & $ 4.7\pm0.3$ & $1.4\pm0.5$ & $0.8\pm0.4$  & $ 1.7\pm0.3$ & $0.9\pm0.5$  & 8773.422$^1$ \\
 $Q(10)$& $0.5\pm0.4$ & $ 4.5\pm0.3$ & $1.0\pm0.5$ & $0.9\pm0.4$  & $ 0.8\pm0.3$ & $2.2\pm0.5u$ & 8780.141 \\
 $P(6)$ & $0.6\pm0.5$ & $ 5.1\pm0.3$ & $1.0\pm0.4$ &              & $ 0.9\pm0.8$ & $1.7\pm0.5u$ & 8782.308 \\
 $Q(12)$&             & $ 2.3\pm0.3$ & $0.6\pm0.4$ &              & $ 1.0\pm0.3$ &              & 8788.558 \\
 $P(8)$ &     	      & $ 2.4\pm0.4$ &             &              & $ 0.8\pm0.3$ &              & 8792.649 \\
 $Q(14)$&             & $ 2.1\pm0.3$ &             &              & $ 0.6\pm0.4$ & $0.6\pm0.3u$ & 8798.459 \\
 $P(10)$&             & $ 2.2\pm0.3$ &             &              &              &              & 8804.499 \\
 $Q(16)$&      	      & $ 2.2\pm0.3$ & $0.8\pm0.4$ &              & $ 0.6\pm0.4$ &              & 8809.842 \\
 $P(12)$&             & $ 0.8\pm0.4$ &             &              &              &              & 8817.827 \\
 $Q(18)$&             & $ 1.8\pm0.5u$&             &              &              &              & 8822.725 \\
 $Q(20)$&             & $ 2.2\pm0.7u$&             &              &              &              & 8837.119 \\
 $Q(22)$&             & $ 0.7\pm0.4u$&             &              &              &              & 8853.041 \\
 $Q(26)$&             & $ 0.7\pm0.5u$&             &              &              &              & 8889.532 \\
\hline
\end{tabular}
\end{table*}
\begin{table*}
\centering
\caption{The same as in Table 2. but for the (3,0) Phillips band.}
\label{Table 4.}
\begin{tabular}{@{}cccccccc}
\hline
$B(N''=J'')$& HD76341 & HD147889    & HD148184 & HD163800 & HD169454 & HD179406 &$\lambda$ [{\AA}]\\
\hline   
$R(6)$ & $0.7\pm0.5u$& $ 1.8\pm0.4$ & $1.0\pm0.4$  &             & $2.9\pm0.6$ & $1.3\pm0.6u$ & 7714.575 \\
$R(4)$ &             & $ 3.5\pm0.4$ & $1.5\pm0.5$  & $0.5\pm0.4$ & $2.0\pm0.5$ & $3.8\pm1.2u$ & 7714.944 \\ 
$R(8)$ & $0.3\pm0.3$ & $ 1.9\pm0.5$ & $0.4\pm0.4u$ &             & $1.0\pm0.5u$&              & 7715.415 \\
$R(2)$ &             & $ 3.7\pm0.4$ & $1.0\pm0.5$  & $0.9\pm0.4u$& $3.5\pm0.5$ & $2.3\pm0.4$  & 7716.528 \\
$R(10)$&             & $ 0.9\pm0.4$ &              &             & $0.6\pm0.5u$& $1.2\pm0.4u$ & 7717.469 \\
$R(0)$ &             & $ 3.5\pm0.4$ & $0.7\pm0.5u$ &             & $3.5\pm0.4$ & $1.7\pm0.5$  & 7719.329 \\
$R(12)$&             & $ 1.0\pm0.4$ & $0.4\pm0.4u$ &             &             &              & 7720.748 \\
$Q(2)$ & $0.7\pm0.3u$& $ 4.8\pm0.4$ & $2.5\pm0.7$  & $1.8\pm0.6$ & $5.9\pm0.5$ & $1.7\pm0.9u$ & 7722.095 \\
$Q(4)$ & $0.7\pm0.4u$& $ 4.5\pm0.4$ & $1.0\pm0.5$  & $2.7\pm0.9$ & $3.2\pm0.4$ & $2.8\pm0.9u$ & 7724.219 \\
$R(14)$&      	     & $ 0.5\pm0.5$ &              &             &             &              & 7725.240 \\
$P(2)$ &             &      	    &              &             & $2.0\pm0.8$ &              & 7725.819 \\ 
$Q(6)$ &             & $ 4.0\pm0.4$ &              &             & $1.4\pm0.6$ & $2.3\pm0.5u$ & 7727.557 \\
$R(16)$&             &      	    &              &             &             &              & 7730.963 \\
$P(4)$ &             & $ 2.2\pm0.6$ & $0.4\pm0.4u$ &             & $0.8\pm0.5$ &              & 7731.663 \\
$Q(8)$ &             & $ 2.4\pm0.6$ &              & $2.1\pm0.6$ &             & $1.1\pm0.4$  & 7732.117 \\
$Q(10)$&      	     & $ 1.9\pm0.4$ &              &             & $0.9\pm0.5$ &              & 7737.904 \\
$P(6)$ &      	     & $ 1.8\pm0.4$ &              &             & $0.4\pm0.4u$&              & 7738.737 \\
$Q(12)$&      	     &      	    &              &             & $1.0\pm0.5$ & $0.7\pm0.7u$ & 7744.900 \\
$P(8)$ &      	     &      	    & $0.3\pm0.3$  &             & $0.8\pm0.5u$&              & 7747.037 \\
$Q(14)$&      	     &      	    &              &             &             &              & 7753.141 \\
$P(10)$&      	     & $ 0.5\pm0.4$ & $0.5\pm0.5$  &             &             &              & 7756.582 \\
$Q(16)$&      	     &      	    &              &             & $0.6\pm0.6u$&              & 7762.623 \\
$P(12)$&      	     &      	    &              &             &             &              & 7767.369 \\
\hline
\end{tabular}
\end{table*}

\begin{table}
\centering
\caption{The same as in Table 2. but for the (4,0) Phillips band.}
\label{Table 5.}
\begin{tabular}{@{}ccc}
\hline
$B(N''=J'')$& HD147889 &$\lambda$[{\AA}]\\
\hline  
$R(4)^a$ & $0.5\pm0.5u$ & 6909.412 \\ 
$R(2)$ & $3.3\pm1.4u$ & 6910.577 \\
$Q(2)$ & $2.0\pm1.0u$ & 6914.975 \\
$Q(4)$ & $1.2\pm0.8u$ & 6916.788 \\
$Q(6)$ & $1.4\pm0.8u$ & 6919.656 \\
\hline
\multicolumn{3}{l}{$^a$ R(4) may form a blend with R(6) 6909.374\,\AA}\\
\end{tabular}
\end{table}
\begin{figure*}
\centerline{
\hbox{
\includegraphics[width=4.8cm,angle=90]{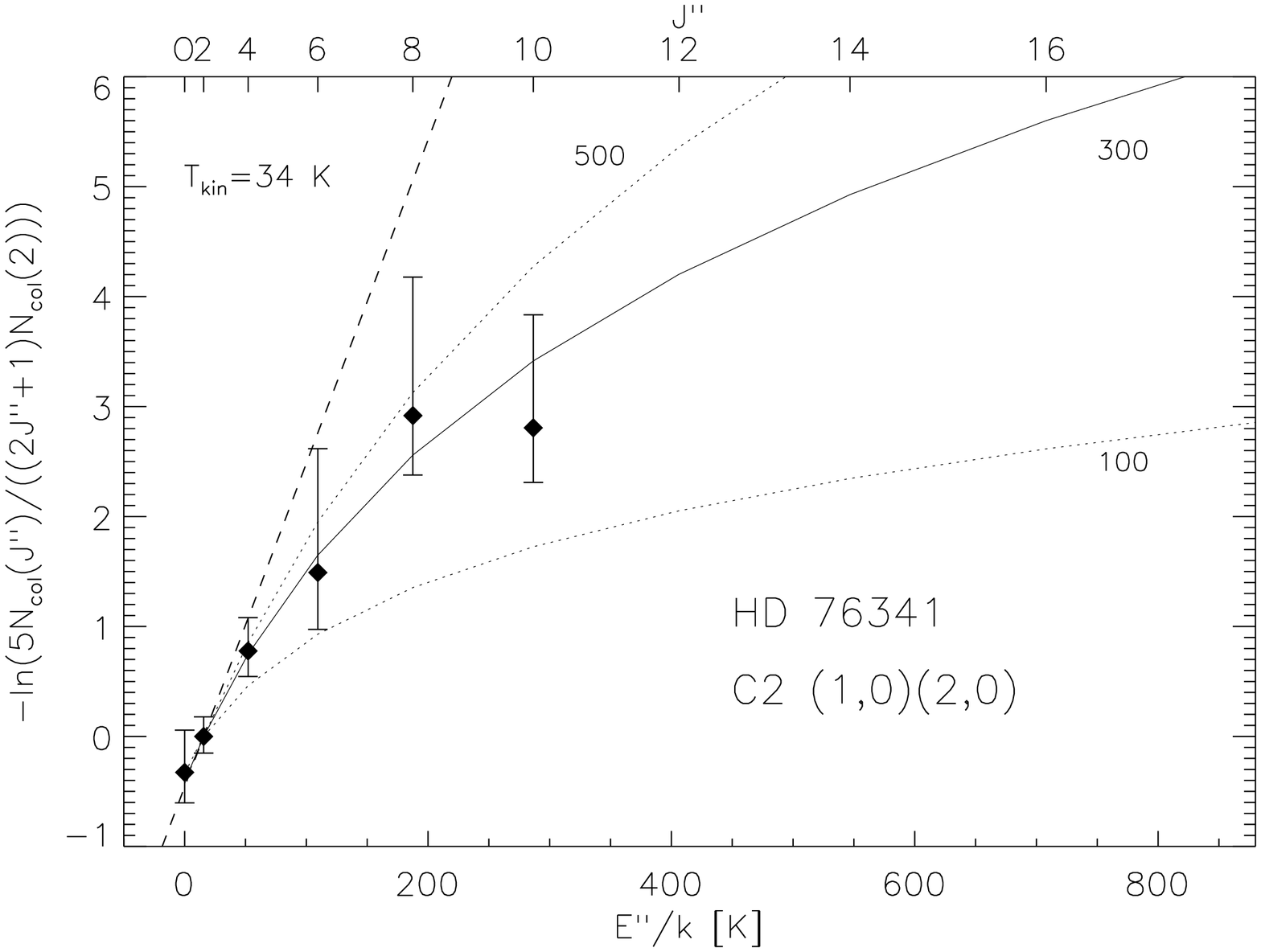}
\includegraphics[width=4.8cm,angle=90]{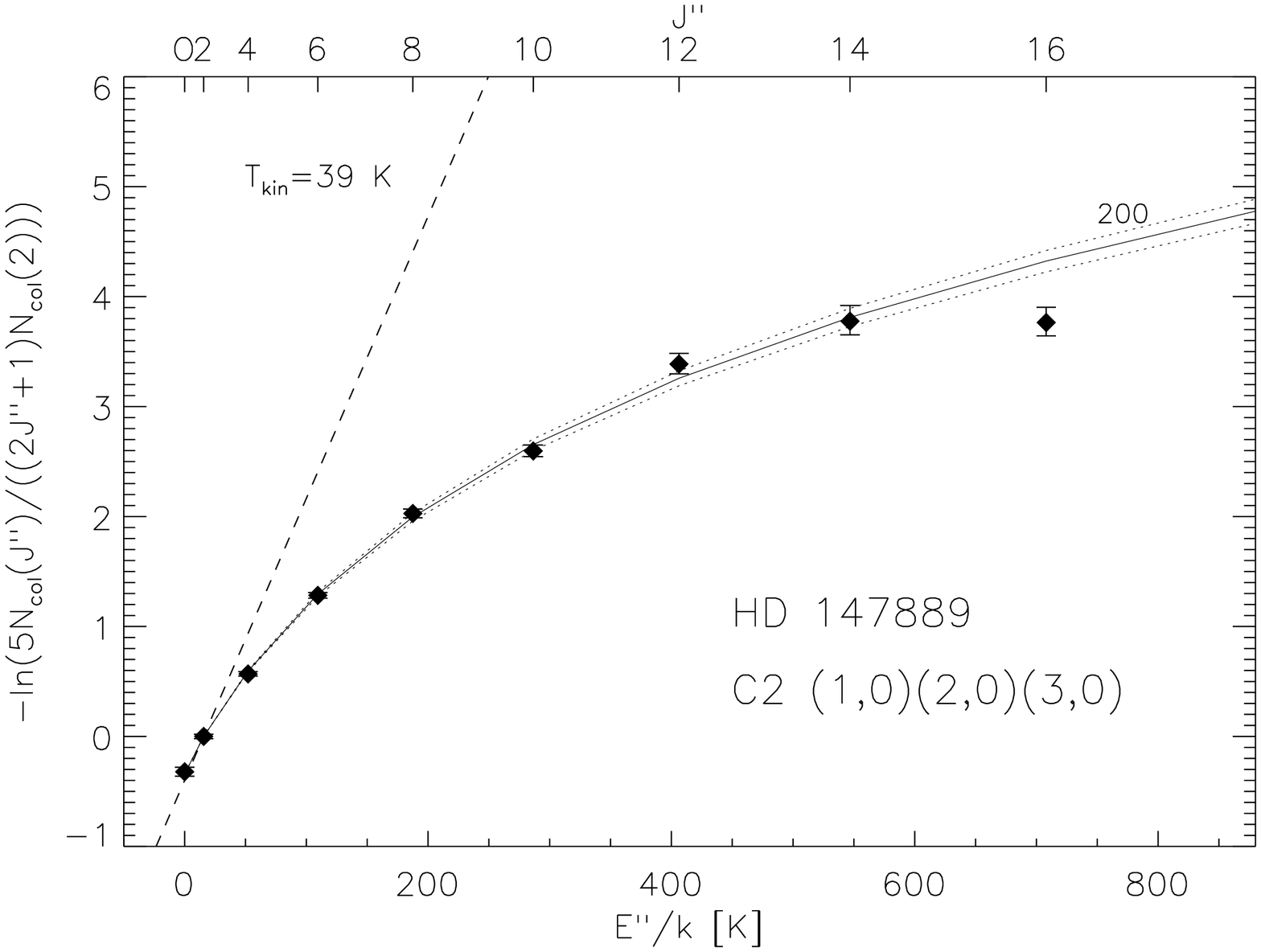}}}
\centerline{
\hbox{
\includegraphics[width=4.8cm,angle=90]{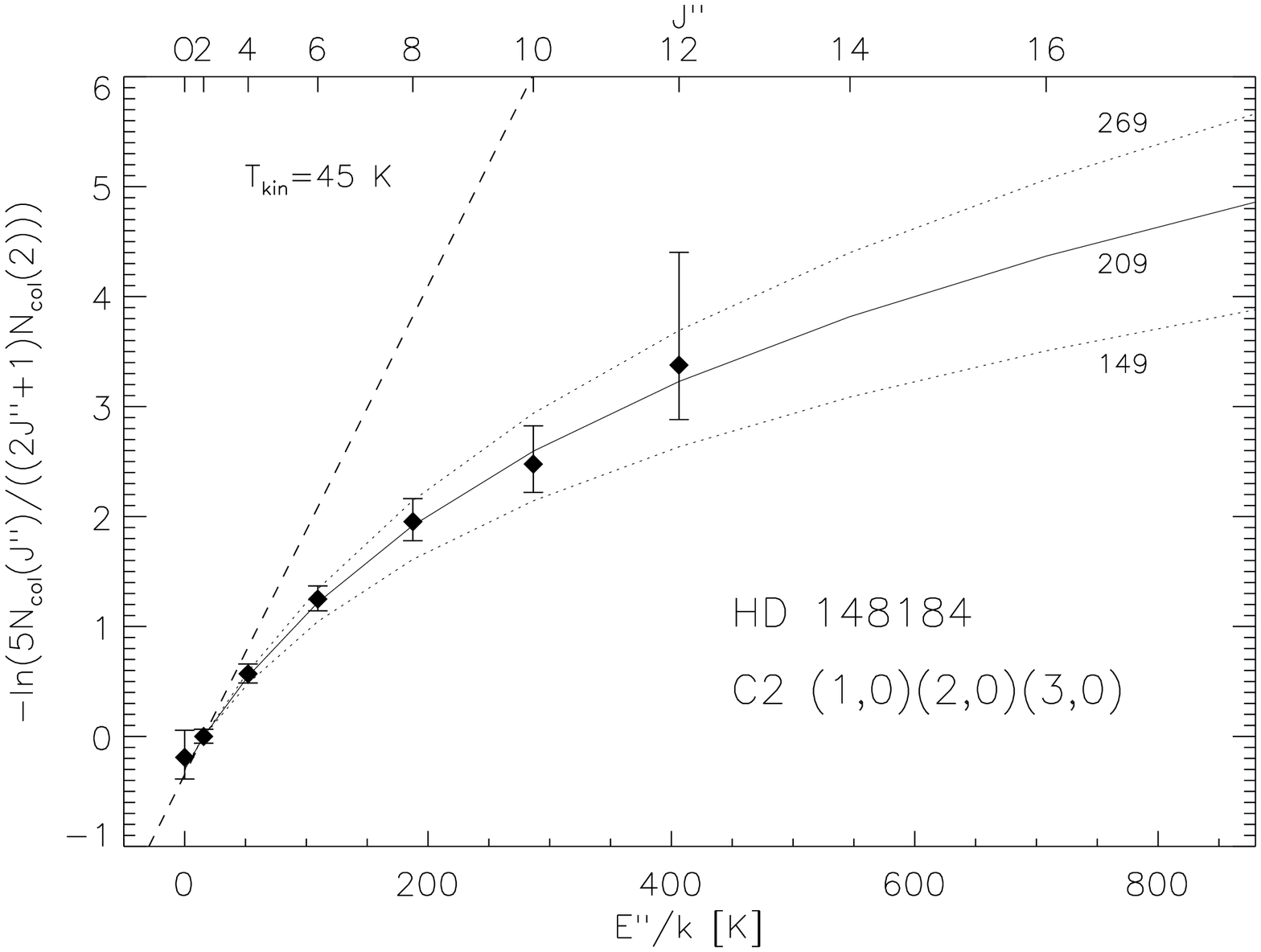}
\includegraphics[width=4.8cm,angle=90]{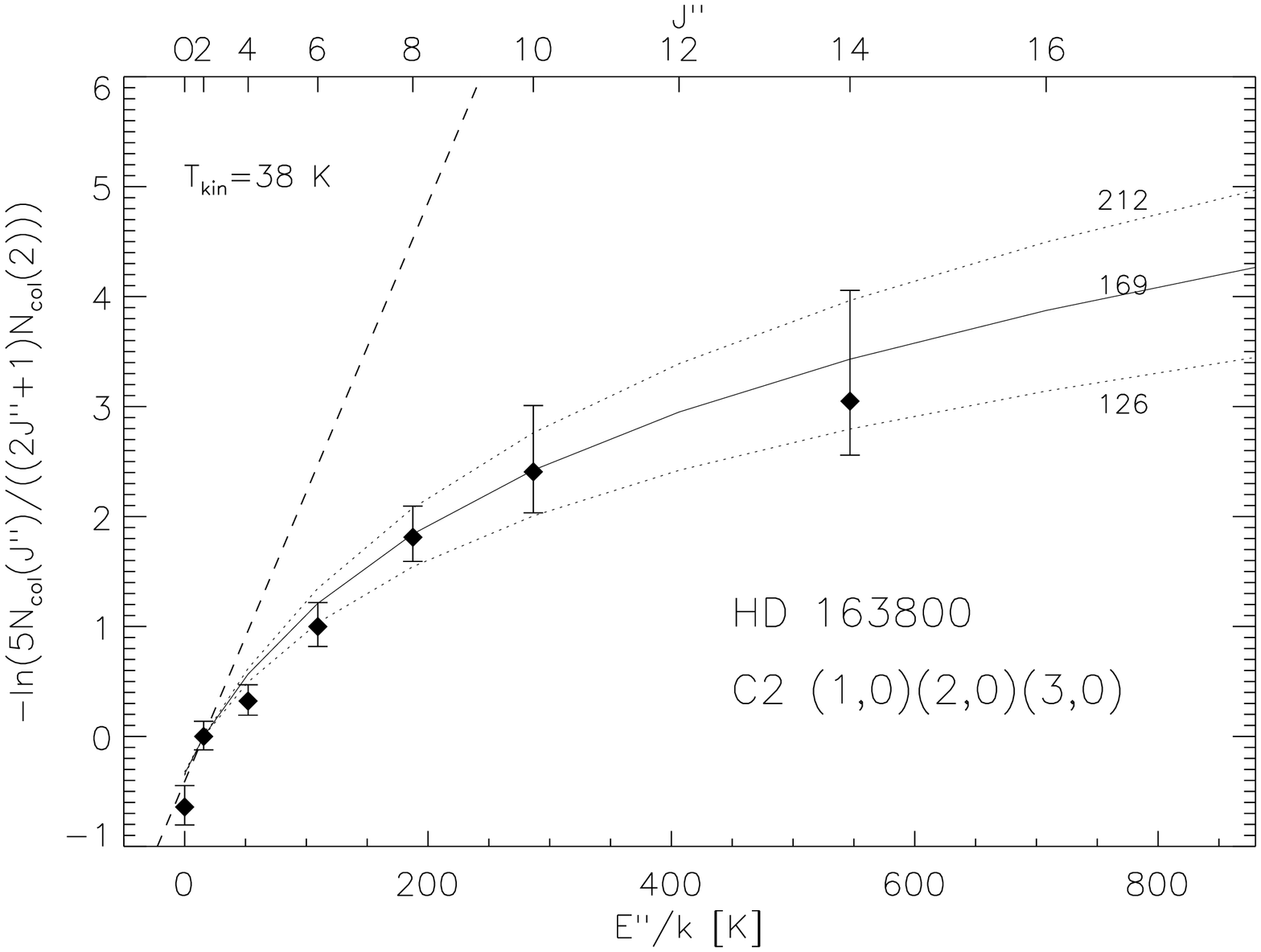}}}
\centerline{
\hbox{
\includegraphics[width=4.8cm,angle=90]{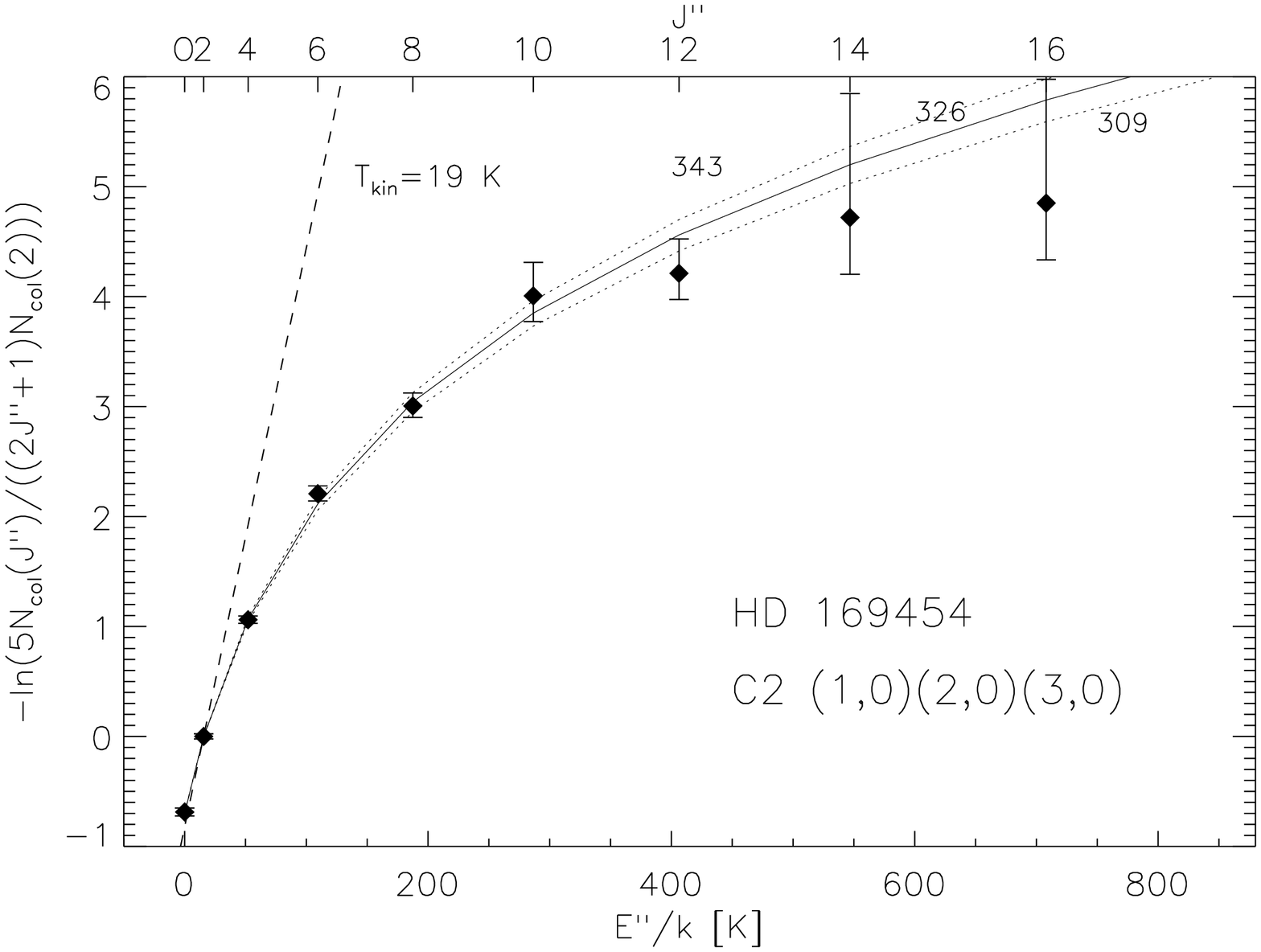}
\includegraphics[width=4.8cm,angle=90]{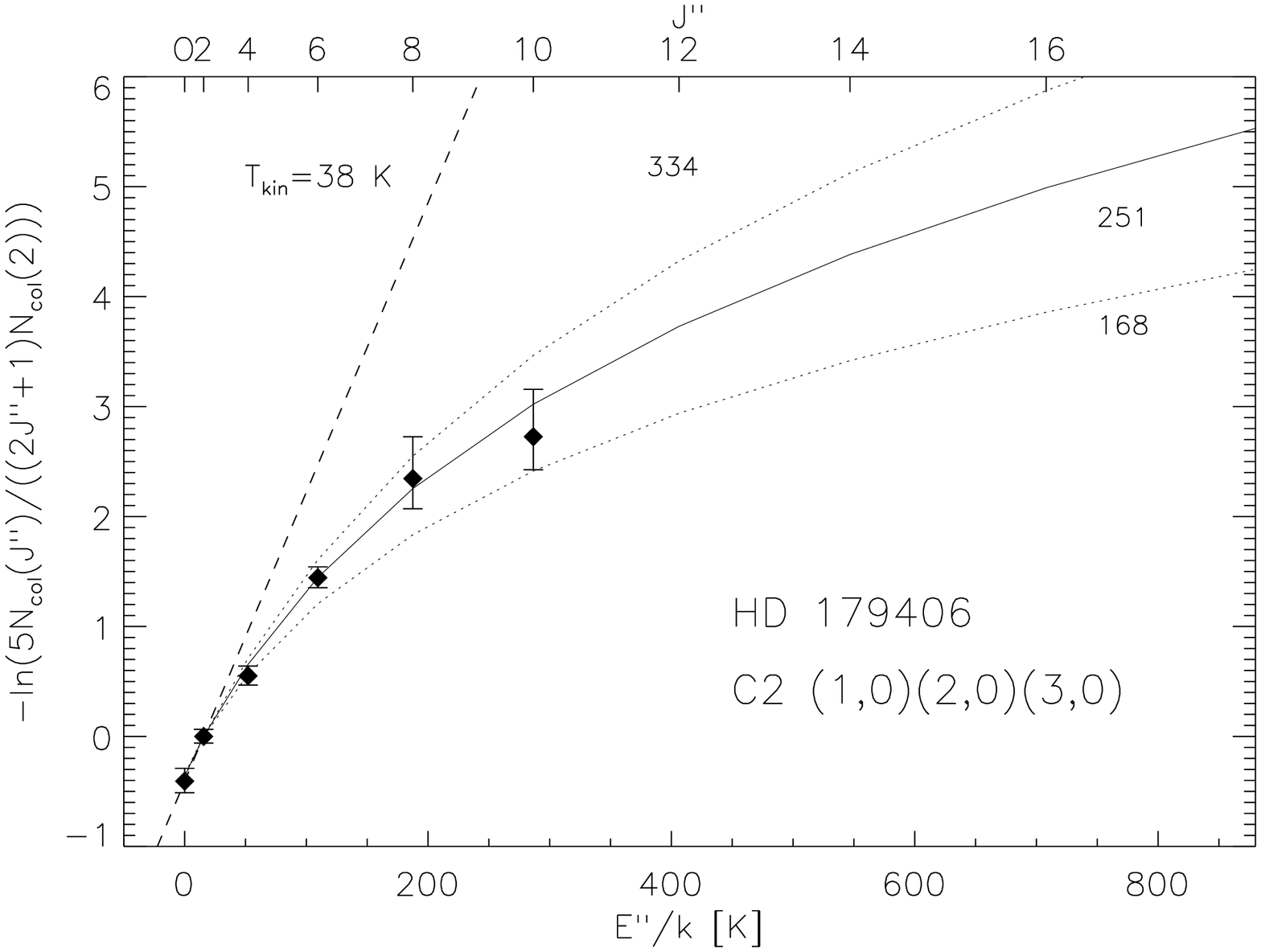}}}
\caption{Relative $C_2$ rotational population diagrams toward six program stars, as a function of the excitation energy (or rotational quantum number $J''$). The solid lines represent fit to the theoretical model, based on the analysis of van Dishoeck \& Black (1982). The straight dashed line shows the best-fitting $T_{02}$.}
\label{fig2}
\end{figure*}
\begin{table*}
\begin{minipage}{150mm}
\centering
\caption{$C_2$ column densities $N_{col}(J'')~[10^{12}~cm^{-2}]$ for each low rotational level $J''$, (N - number of measurements used to determined given $N_{col}$).}
\label{Table 6.}
\begin{tabular}{@{}lclclclclclcl}
\hline
   & HD\,76341   &   & HD\,147889   &   & HD\,148184  &   & HD\,163800  &   & HD\,169454& & HD\,179406 &\\
\hline
$J''$&$N_{col}$&N&$N_{col}$&N&$N_{col}$&N&$N_{col}$&N&$N_{col}$&N&$N_{col}$&N\\
\hline
0  & $0.9\pm0.3$ & 2 & $ 8.5\pm0.4$ & 3 & $1.6\pm0.3$ & 2 & $1.9\pm0.3$ & 2 & $ 9.5\pm0.4$ &2& $ 3.1\pm0.4$ & 2 \\
2  & $3.4\pm0.6$ & 3 & $30.7\pm0.6$ & 8 & $6.7\pm0.4$ & 5 & $4.9\pm0.6$ & 6 & $23.8\pm0.6$ &6& $10.4\pm0.7$ & 6 \\
4  & $2.8\pm0.7$ & 3 & $31.3\pm0.6$ & 9 & $6.8\pm0.6$ & 8 & $6.4\pm0.9$ & 6 & $14.8\pm0.5$ &6& $10.8\pm1.0$ & 4 \\
6  & $2.0\pm1.4$ & 1 & $22.1\pm0.6$ & 9 & $5.0\pm0.6$ & 7 & $4.7\pm0.9$ & 2 & $ 6.8\pm0.5$ &6& $ 6.4\pm0.6$ & 4 \\
8  & $0.6\pm0.5$ & 2 & $13.7\pm0.5$ & 8 & $3.2\pm0.6$ & 6 & $2.7\pm0.7$ & 4 & $ 4.0\pm0.5$ &6& $ 3.4\pm1.1$ & 2 \\
10 & $0.9\pm0.6$ & 2 & $ 9.6\pm0.5$ & 9 & $2.3\pm0.7$ & 4 & $1.9\pm0.8$ & 1 & $ 1.8\pm0.5$ &3& $ 2.9\pm1.0$ & 1 \\
12 &             &   & $ 5.2\pm0.5$ & 6 & $1.1\pm0.7$ & 2 &             &   & $ 1.8\pm0.5$ &3&              &   \\
14 &             &   & $ 4.1\pm0.6$ & 4 &             &   & $1.3\pm0.9$ & 1 & $ 1.2\pm0.8$ &1&              &   \\
16 &             &   & $ 4.7\pm0.6$ & 2 &             &   &             &   & $ 1.2\pm0.8$ &1&              &   \\
\hline
\end{tabular}
\end{minipage}
\end{table*}
\begin{table*}
\begin{minipage}{150mm}
\centering
\caption{Summary of the observational data for $C_2$, $T_{02}$ ($T_{04}$, $T_{06}$) - rotational temperature calculated from the two (three, four) lowest rotational levels, $N_{col}$ - total column densities and the results of a model: $T_{kin}$ - gas kinetic temperature, $n_c$ - the effective density of collision partners and $N_{col}(J''=2)$ - column density derived from $J''=2$.}
\label{Table 7.}
\begin{tabular}{@{}llllc|ccc}
\hline
object & $T_{02} [K]$ & $T_{04} [K]$ & $T_{06} [K]$ & $N_{col}
[10^{12}\rmn{cm}^{-2}]$ & $T_{kin} [K]$&$n_c [\rmn{cm}^{-3}]$& $N_{col}(J''=2)
[10^{12}\rmn{cm}^{-2}]$\\
\hline
HD\,76341  & $48\pm48$  & $47\pm16$ & $52\pm15$ & $ 11\pm1$ & $34\pm20$ & $300\pm200$ & $ 3.3\pm0.5$ \\
HD\,147889 & $49\pm7$  & $62\pm 3$ & $71\pm 2$ & $133\pm1$ & $39\pm2$ & $199\pm7$ & $30.8\pm0.5$ \\
HD\,148184 & $82\pm82$ & $65\pm 12$ & $74\pm7$& $ 30\pm3$ & $45\pm12$ & $209\pm60$ & $6.6\pm0.4$\\
HD\,163800&$24\pm  8$  &$66\pm 17$&$76\pm13$  & $ 28\pm1$ & $38\pm15$&$169\pm43$ &$ 5.6\pm0.5$\\
HD\,169454&$23\pm  2$  &$31\pm  1$&$36\pm 1$  & $ 65\pm1$ & $19\pm2$ &$326\pm17$ &$23.8\pm0.5$\\
HD\,179406 & $38\pm12$ & $59\pm9$ & $62\pm 5$ & $ 39\pm1$ & $38\pm9$ & $251\pm83$ & $10.8\pm0.5$ \\
\hline
\end{tabular}
\end{minipage}
\end{table*} 

For the interpretation of the rotational diagrams we have constructed a grid
of models based on the radiative excitation model of \cite{Dishoeck1982}.
The detailed behaviour of excitation temperature depends on the gas kinetic
temperature and on the ratio of the collisional rate $n_c \times \sigma$ to
intensity of the average galactic field, here expressed by $I$. The parameter 
$I$ has been introduced by \cite{Dishoeck1982} as a scaling factor of the 
standard field adopted in their paper.
With the lack of detailed information on the
radiation field in individual diffuse clouds we made a crude assumption
$I = 1$. Then, from the best fitting models of excitation, one can estimate
the density of collision partners in a molecular cloud~$n_c$. 
Following \cite{Dishoeck1982},
we computed grid of models with step of $5\,\rmn{K}$ in kinetic temperature and
25\,cm$^{-3}$ in collisional partner densities. The models were constructed
by solving eqn. 27 in \cite{Dishoeck1982}, using their quadrupole and
collisional rates and radiative excitation matrix.
To adjust for the
differences in adopted f-values of Phillips transitions we rescaled the
density of collisional partners according to \cite{Dishoeck1984}
prescription by factor 1.59 (factor of 1.35 from \cite{Dishoeck1984},
times 1.18 for difference in f-values used here).
The models with original f-values of \cite{Dishoeck1982} 
were checked to reproduce results of original paper
with reliable accuracy of 20 percent in the worst case of the higher levels.
The grid of models was used to find the best fit to the
earlier determined column densities individually for each transitions 
weighted by their errors.
We decided to search for the best fit to the absolute column densities,
instead of the relative to $J''=2$ populations. Hence, one additional parameter 
of the model was absolute column density $N(J''= 2)$ of $J''=2$ level
(which is equivalent to the total column density of $C_2$).
This approach is still consistent with the fact that kinetic temperature
and density of collisional partners depends only on the relative
populations but changes the way how observational errors are propagated
in the process of finding the best fit model.
In result, three independent parameters: gas kinetic temperature
($T_{kin}$), collisional partners density $n_c$, 
and column density of the $J''=2$ level were estimated simultaneously 
for the set of observed column densities (see Table 7).
The best fitted models are presented on the rotational diagrams (see
Figure~2, the labels describe appropriate values of $n_c$).
The analogous procedure applied for the relative populations gives
very similar results for $T_{kin}$ and $n_c$, well inside determined errors.
The difference in kinetic temperatures determined from both methods is less 
than 1\,K except HD\,148184 were it is less than 3\,K.
%
Note that because of nonlinearity of the model the errors of best fitted
parameters may be very asymmetric. Especially in these cases when error 
is a significant fraction of fitted values, e.g. HD\,76341 and HD\,163800,
the uncertainty of $T_{kin}$ toward lower values will be less than
the uncertainty toward higher values. In consequence $n_c$ could be much
higher toward both objects.

We also derived a set of rotational temperatures: $T_{02}$, $T_{04}$,
$T_{06}$ corresponding to the mean excitation temperatures derived from
a linear fit to logarithm of column densities of the
first: two, three and four levels, respectively, starting from 
\mbox{$J''= 0$} respectively.
$T_{02}$ is the best estimator of the gas kinetic temperature, 
$T_{02} \geq T_{kin}$, but generally it is not well determined, because
there is only one available line R(0) in each band absorbing from the $J''=0$ level.
$T_{04}$ was more often published than $T_{06}$, but the latter is usually determined with
better precision. On the other side, both $T_{04}$ and $T_{06}$ depend
on collisional density and radiation intensity in increasing level.
Table 7 summarises all these parameters \footnote{
Excitation temperatures determined in this paper are slightly different from
those announced in the final version of our paper \cite{Kazmierczak2009}.
The reason of differences was reanalysis of errors of equivalent widths. 
}.

\begin{figure}
\begin{center}
\includegraphics[width=4.5cm,height=6cm]{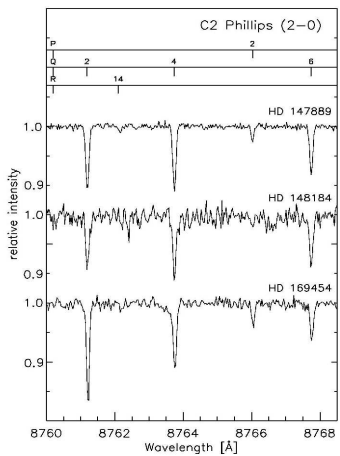}
\caption{
The comparison of intensities of few lines of the (2,0) band for 3 objects;
the spectra were normalised to the line Q(4) 8763.74 \AA~(2,0) of
HD\,147889. The excitation temperatures $T_{06}$ of $C_2$ in HD\,148184,
HD\,147889 and HD\,169454 amount 74, 71 and 36\,K, respectively.}
\label{fig3}
\end{center}
\end{figure}

\section{Discussion}
The $C_2$ molecule in interstellar clouds was observed many times, but
usually measurements were limited to the most easily available to
ground-based instruments lines of the (2,0) Phillips band. Apart from that
measurements have usually been done with spectrographs of moderate
resolution and signal-to-noise ratio
\citep{Snow1978,Hobbs1979,Danks1983,Dishoeck1984,Gredel1986,Dishoeck1989,Federman1994}.

We measured absorption lines of the interstellar $C_2$ toward six reddened
early type stars. The presence of $C_2$ in the sight line toward HD\,76341
and HD\,163800 has not been reported before.

Generally, excitation temperatures $T_{04}$ in our program stars tend to be
above 47\,K. One evident exception is HD\,169454 (31\,K). This cloud is
definitely different from the others. There is the lowest gas kinetic 
temperature. Van Dishoeck \& Black (1989) suggested that it is a real translucent cloud,
whereas the other clouds are only diffuse interstellar clouds. 
 
The behaviour of excitation temperatures may be seen directly by
inspection of the portions of spectra around the Q(J''=4) transition of the
(2,0) band (see Fig. 3). Note, that the spectrum was renormalized to the
strength of Q(4) absorption. It is easily seen, that 
excitation temperature in HD\,147889 is almost identical to that
of HD\,148184 and higher than the excitation temperature of HD\,169454,
which is confirmed in the quantitative analysis presented above.

Simultaneous observations of different vibrational bands give possibility to
compare individually determined column densities for each band. For this
purpose data collected for HD\,147889 are especially useful.
The observed column densities determined individually from bands 
(1,0), (2,0) and (3,0) amount
\mbox{$(11.2 \pm 0.4) \times 10^{13} \rmn{cm}^{-3}$}, 
\mbox{$(13.3 \pm 0.2) \times 10^{13} \rmn{cm}^{-3}$}, and 
\mbox{$(13.2 \pm 0.8) \times 10^{13} \rmn{cm}^{-3}$} respectively.
The observed column density determined form all the bands
together amounts \mbox{$(13.0 \pm 0.2) \times 10^{13} \rmn{cm}^{-3}$.}
In conclusion, the determination of column
densities from only one of these bands gives reasonable value of the $C_2$
column density; but the possibility of analysing lines of many different
bands enhances the accuracy of the results. It excludes accidental errors
(e.g. cosmic rays, telluric lines, instrumental errors) which are impossible
to remove when only one band is available.

The interstellar absorption
lines of $C_2$ toward HD\,163800 arise in a single velocity component at a radial velocity,
$V_{LSR}$ = 7.2 km\,s$^{-1}$ with respect to the local standard of rest. This radial
velocity and position of the star suggest an association of the intervening
molecular cloud with a large cloud of cold atomic hydrogen first surveyed by
\cite{Riegel1972} (Riegel-Crutcher cloud) as a prominent self-absorption feature in the
21 cm line. The distance to the cold atomic cloud is constrained by
observations of optical lines in direction to background stars to be $125
\pm 25$ pc \citep{Crutcher1984}. Apparent association with atomic cloud does
not imply physical connection with molecular cloud responsible for the
observed absorptions of $C_2$. Small molecular cloud in the direction to
HD\,169454 is also possibly associated with the Riegel-Crutcher HI
cloud; it was analysed by \cite{Jannuzi1988}. The distance to HD\,163800
from the spectroscopic data is estimated to be about 1.5 kpc.

The detailed comparison between our results and these of previous papers is
shown in Appendix.

\section{Summary}

From the inspection of absorption lines in direction to HD\,147889, we have
found that individual Phillips bands (1,0), (2,0) and (3,0) give consistent
results, both for column densities and excitation temperatures. However not
every feature is the same easy to measure in all bands; e.g. the P(4) and Q(8)
transitions are the easiest to separate in the (1,0) band and lines of R
branch in (3,0) one (R-branch lines of the (2,0) band very often are located
near to or in the wings of the stellar line HI what makes measurements
uncertain). In this paper measurements of three Phillips bands were used to
improve reliability of results due to growing number of identified lines and
their separation.

The assumption of the identical radiation field for all clouds allows to
calculate the density of collision partners $n_c$; the found values range
from about 170 to 330 $\rmn{cm}^{-3}$ (100 to 500 if errors are included). 
Absolute values may be different because of
the peculiar radiation field and approximate treatment of the collisional
excitation rate of $C_2$ with $H_2$. If the rotational population
distributions are useful as diagnostic tools of the cloud densities, it is
essential to have better understanding of cross sections of rotationally
inelastic collisions of $C_2$ with $H_2$. Determination of $n_c$ is less
dependent on the accurate value of the gas kinetic temperature. 
The visible transitions of $C_2$ are relatively easily observed,
(in contrast to the far-UV bands of $H_2$ and
very weak lines of $C_3$), and provide
an important tool for the determination of
the physical and chemical conditions of diffuse clouds.

The presence of $C_2$ in the measurable amount is reported for the two new
lines of sight: toward HD\,76341 and HD\,163800. The latter star lies in the
background of the Riegel-Crutcher cloud of the cold HI. The molecular cloud,
origin of $C_2$ absorptions, may be physically associated with the atomic
hydrogen cloud, similarly to the earlier considered small molecular cloud in
direction to HD\,169454 \citep{Jannuzi1988}.

To sum up, the high resolution and signal-to-noise ratio spectra acquired
with the ESO instrument allow us to study densities and rotational
temperatures varying from object to object with increasingly accuracy.
This is seen particularly well when compare the theoretical model
with the observational data (e.g. HD~147889 in Figure 2).

We are planning to continue the survey of high resolution and high
signal-to-noise ratio spectra with detected interstellar diatomic carbon and
compare with other interstellar absorption features. It is interesting
whether other molecules are spatially correlated with $C_2$; this may
efficiently constrain interstellar chemistry.

\section*{Acknowledgments}

We thank Daniel Welty for a review and significat comments that improved the
manuscript.
This paper was supported from the grants: N203 019 31/2874, N203 012 32/1550
and N203 393334 of the Science and High Education Ministry of Poland.

\bibliographystyle{mn2e}
\bibliography{ms}

\begin{thebibliography}{}

\bibitem[\protect\citeauthoryear{{Bagnulo}, {Jehin}, {Ledoux}, {Cabanac},
  {Melo}, {Gilmozzi} \& {The ESO Paranal Science Operations Team}}{{Bagnulo}
  et~al.}{2003}]{Bagnulo2003}
{Bagnulo} S.,  {Jehin} E.,  {Ledoux} C.,  {Cabanac} R.,  {Melo} C.,  {Gilmozzi}
  R.,    {The ESO Paranal Science Operations Team} 2003, The Messenger, 114, 10

\bibitem[\protect\citeauthoryear{{Bakker}, {van Dishoeck}, {Waters} \&
  {Schoenmaker}}{{Bakker} et~al.}{1997}]{Bakker1997}
{Bakker} E.~J.,  {van Dishoeck} E.~F.,  {Waters} L.~B.~F.~M.,    {Schoenmaker}
  T.,  1997, A\&A, 323, 469

\bibitem[\protect\citeauthoryear{{Bakker}, {Waters}, {Lamers}, {Trams} \& {van
  der Wolf}}{{Bakker} et~al.}{1996}]{Bakker1996}
{Bakker} E.~J.,  {Waters} L.~B.~F.~M.,  {Lamers} H.~J.~G.~L.~M.,  {Trams}
  N.~R.,    {van der Wolf} F.~L.~A.,  1996, A\&A, 310, 893

\bibitem[\protect\citeauthoryear{{Ballik} \& {Ramsay}}{{Ballik} \&
  {Ramsay}}{1963}]{BallikRamsay1963}
{Ballik} E.~A.,  {Ramsay} D.~A.,  1963, ApJ, 137, 61

\bibitem[\protect\citeauthoryear{{Ballik} \& {Ramsay}}{{Ballik} \&
  {Ramsay}}{1963}]{BallikRamsay1963b}
{Ballik} E.~A.,  {Ramsay} D.~A.,  1963, ApJ, 137, 64

\bibitem[\protect\citeauthoryear{{Black} \& {Dalgarno}}{{Black} \&
  {Dalgarno}}{1977}]{Black1977}
{Black} J.~H.,  {Dalgarno} A.,  1977, ApJS, 34, 405

\bibitem[\protect\citeauthoryear{{Black}, {Hartquist} \& {Dalgarno}}{{Black}
  et~al.}{1978}]{Black1978}
{Black} J.~H.,  {Hartquist} T.~W.,    {Dalgarno} A.,  1978, ApJ, 224, 448

\bibitem[\protect\citeauthoryear{{Bondar}, {Kozak}, {Gnaci{\'n}ski},
  {Galazutdinov}, {Beletsky} \& {Kre{\l}owski}}{{Bondar}
  et~al.}{2007}]{Bondar2007}
{Bondar} A.,  {Kozak} M.,  {Gnaci{\'n}ski} P.,  {Galazutdinov} G.~A.,
  {Beletsky} Y.,    {Kre{\l}owski} J.,  2007, MNRAS, 378, 893



\bibitem[\protect\citeauthoryear{{Chaffee} Jr. \& {Lutz}}{{Chaffee} \&
  {Lutz}}{1978}]{Chaffee1978}
{Chaffee} Jr. F.~H.,  {Lutz} B.~L.,  1978, ApJ, 221, L91

\bibitem[\protect\citeauthoryear{{Chaffee} Jr., {Lutz}, {Black}, {Vanden Bout}
  \& {Snell}}{{Chaffee} et~al.}{1980}]{Chaffee1980}
{Chaffee} Jr. F.~H.,  {Lutz} B.~L.,  {Black} J.~H.,  {Vanden Bout} P.~A.,
  {Snell} R.~L.,  1980, ApJ, 236, 474

\bibitem[\protect\citeauthoryear{{Chauville}, {Maillard} \&
  {Mantz}}{{Chauville} et~al.}{1977}]{Chauville1977}
{Chauville} J.,  {Maillard} J.~P.,    {Mantz} A.~W.,  1977, Journal of
  Molecular Spectroscopy, 68, 399

\bibitem[\protect\citeauthoryear{{Crawford}}{{Crawford}}{1997}]{Crawford1997}
{Crawford} I.~A.,  1997, MNRAS, 290, 41

\bibitem[\protect\citeauthoryear{{Crawford} \& {Barlow}}{{Crawford} \&
  {Barlow}}{1996}]{Crawford1996}
{Crawford} I.~A.,  {Barlow} M.~J.,  1996, MNRAS, 280, 863

\bibitem[\protect\citeauthoryear{{Crutcher}}{{Crutcher}}{1985}]{Crutcher1985}
{Crutcher} R.~M.,  1985, ApJ, 288, 604

\bibitem[\protect\citeauthoryear{{Crutcher} \& {Chu}}{{Crutcher} \&
  {Chu}}{1985}]{Crutcher_Chu1985}
{Crutcher} R.~M.,  {Chu} Y.-H.,  1985, ApJ, 290, 251

\bibitem[\protect\citeauthoryear{{Crutcher} \& {Lien}}{{Crutcher} \&
  {Lien}}{1984a}]{Crutcher1984}
{Crutcher} R.~M.,  {Lien} D.~J.,  1984a, Technical report, {Distances of local
  clouds from optical line observations}



\bibitem[\protect\citeauthoryear{{Danks} \& {Lambert}}{{Danks} \&
  {Lambert}}{1983}]{Danks1983}
{Danks} A.~C.,  {Lambert} D.~L.,  1983, A\&A, 124, 188

\bibitem[\protect\citeauthoryear{{Dekker}, {D'Odorico}, {Kaufer}, {Delabre} \&
  {Kotzlowski}}{{Dekker} et~al.}{2000}]{uves}
{Dekker} H.,  {D'Odorico} S.,  {Kaufer} A.,  {Delabre} B.,    {Kotzlowski} H.,
  2000, in {M.~Iye \& A.~F.~Moorwood} ed., Society of Photo-Optical
  Instrumentation Engineers (SPIE) Conference Series Vol.~4008 of Society of
  Photo-Optical Instrumentation Engineers (SPIE) Conference Series, {Design,
  construction, and performance of UVES, the echelle spectrograph for the UT2
  Kueyen Telescope at the ESO Paranal Observatory}.
pp 534--545

\bibitem[\protect\citeauthoryear{{Douay}, {Nietmann} \& {Bernath}}{{Douay}
  et~al.}{1988}]{Douay1988}
{Douay} M.,  {Nietmann} R.,    {Bernath} P.~F.,  1988, Journal of Molecular
  Spectroscopy, 131, 261

\bibitem[\protect\citeauthoryear{{Douglas}}{{Douglas}}{1977}]{Douglas1977}
{Douglas} A.~E.,  1977, Nature, 269, 130

\bibitem[\protect\citeauthoryear{{Erman} \& {Iwamae}}{{Erman} \&
  {Iwamae}}{1995}]{Erman1995}
{Erman} P.,  {Iwamae} A.,  1995, ApJ, 450, L31+

\bibitem[\protect\citeauthoryear{{Federman}, {Strom}, {Lambert}, {Cardelli},
  {Smith} \& {Joseph}}{{Federman} et~al.}{1994}]{Federman1994}
{Federman} S.~R.,  {Strom} C.~J.,  {Lambert} D.~L.,  {Cardelli} J.~A.,  {Smith}
  V.~V.,    {Joseph} C.~L.,  1994, ApJ, 424, 772

\bibitem[\protect\citeauthoryear{{Frisch}}{{Frisch}}{1972}]{Frisch1972}
{Frisch} P.,  1972, ApJ, 173, 301

\bibitem[\protect\citeauthoryear{{Galazutdinov}}{{Galazutdinov}}{1992}]{Galazu%
tdinov1992}
{Galazutdinov} G.,  1992, Preprint Spets. Astrof. Obs. Russian, No. 92

\bibitem[\protect\citeauthoryear{{Gredel}}{{Gredel}}{1999}]{Gredel1999}
{Gredel} R.,  1999, A\&A, 351, 657

\bibitem[\protect\citeauthoryear{{Gredel} \& {Muench}}{{Gredel} \&
  {Muench}}{1986}]{Gredel1986}
{Gredel} R.,  {Muench} G.,  1986, A\&A, 154, 336

\bibitem[\protect\citeauthoryear{{Gredel}, {van Dishoeck} \& {Black}}{{Gredel}
  et~al.}{1989}]{Gredel1989}
{Gredel} R.,  {van Dishoeck} E.~F.,    {Black} J.~H.,  1989, ApJ, 338, 1047

\bibitem[\protect\citeauthoryear{{Gredel}, {van Dishoeck} \& {Black}}{{Gredel}
  et~al.}{1991}]{Gredel1991}
{Gredel} R.,  {van Dishoeck} E.~F.,    {Black} J.~H.,  1991, A\&A, 251, 625

\bibitem[\protect\citeauthoryear{{Grevesse} \& {Sauval}}{{Grevesse} \&
  {Sauval}}{1973}]{Grevesse1973}
{Grevesse} N.,  {Sauval} A.~J.,  1973, A\&A, 27, 29

\bibitem[\protect\citeauthoryear{{Herbig} \& {Leka}}{{Herbig} \&
  {Leka}}{1991}]{Herbig91}
{Herbig} G.~H.,  {Leka} K.~D.,  1991, ApJ, 382, 193

\bibitem[\protect\citeauthoryear{{Hobbs}}{{Hobbs}}{1979}]{Hobbs1979}
{Hobbs} L.~M.,  1979, ApJ, 232, L175

\bibitem[\protect\citeauthoryear{{Hobbs}, {York}, {Snow}, {Oka}, {Thorburn},
  {Bishof}, {Friedman}, {McCall}, {Rachford}, {Sonnentrucker} \&
  {Welty}}{{Hobbs} et~al.}{2008}]{Hobbs2008}
{Hobbs} L.~M.,  {York} D.~G.,  {Snow} T.~P.,  {Oka} T.,  {Thorburn} J.~A.,
  {Bishof} M.,  {Friedman} S.~D.,  {McCall} B.~J.,  {Rachford} B.,
  {Sonnentrucker} P.,    {Welty} D.~E.,  2008, ApJ, 680, 1256

\bibitem[\protect\citeauthoryear{{Hunter}, {Smoker}, {Keenan}, {Ledoux},
  {Jehin}, {Cabanac}, {Melo} \& {Bagnulo}}{{Hunter} et~al.}{2006}]{Hunter2006}
{Hunter} I.,  {Smoker} J.~V.,  {Keenan} F.~P.,  {Ledoux} C.,  {Jehin} E.,
  {Cabanac} R.,  {Melo} C.,    {Bagnulo} S.,  2006, MNRAS, 367, 1478

\bibitem[\protect\citeauthoryear{{Jannuzi}, {Black}, {Lada} \& {van
  Dishoeck}}{{Jannuzi} et~al.}{1988}]{Jannuzi1988}
{Jannuzi} B.~T.,  {Black} J.~H.,  {Lada} C.~J.,    {van Dishoeck} E.~F.,  1988,
  ApJ, 332, 995

\bibitem[\protect\citeauthoryear{{Johnson}, {Fink} \& {Larson}}{{Johnson}
  et~al.}{1983}]{Johnson1983}
{Johnson} J.~R.,  {Fink} U.,    {Larson} H.~P.,  1983, ApJ, 270, 769

\bibitem[\protect\citeauthoryear{{Ka{\'z}mierczak}, {Gnaci{\'n}ski}, {Schmidt},
  {Galazutdinov}, {Bondar} \& {Kre{\l}owski}}{{Ka{\'z}mierczak}
  et~al.}{2009}]{Kazmierczak2009}
{Ka{\'z}mierczak} M.,  {Gnaci{\'n}ski} P.,  {Schmidt} M.~R.,  {Galazutdinov}
  G.,  {Bondar} A.,    {Kre{\l}owski} J.,  2009, A\&A, 498, 785

\bibitem[\protect\citeauthoryear{{Lambert}}{{Lambert}}{1978}]{Lambert1978}
{Lambert} D.~L.,  1978, MNRAS, 182, 249

\bibitem[\protect\citeauthoryear{{Lambert} \& {Danks}}{{Lambert} \&
  {Danks}}{1983}]{Lambert1983}
{Lambert} D.~L.,  {Danks} A.~C.,  1983, ApJ, 268, 428

\bibitem[\protect\citeauthoryear{{Lambert}, {Sheffer} \& {Federman}}{{Lambert}
  et~al.}{1995}]{Lambert1995}
{Lambert} D.~L.,  {Sheffer} Y.,    {Federman} S.~R.,  1995, ApJ, 438, 740

\bibitem[\protect\citeauthoryear{{Langhoff}, {Bauschlicher} Jr., {Rendell} \&
  {Komornicki}}{{Langhoff} et~al.}{1990}]{Langhoff1990}
{Langhoff} S.~R.,  {Bauschlicher} Jr. C.~W.,  {Rendell} A.~P.,    {Komornicki}
  A.,  1990, J. Chem. Phys., 92, 6599

\bibitem[\protect\citeauthoryear{{Lien}}{{Lien}}{1984}]{Lien1984}
{Lien} D.~J.,  1984, ApJ, 287, L95

\bibitem[\protect\citeauthoryear{{Marenin} \& {Johnson}}{{Marenin} \&
  {Johnson}}{1970}]{Marenin1970}
{Marenin} I.~R.,  {Johnson} H.~R.,  1970, Journal of Quantitative Spectroscopy
  and Radiative Transfer, 10, 305


\bibitem[\protect\citeauthoryear{{Mayer} \& {O'dell}}{{Mayer} \&
  {O'dell}}{1968}]{Mayer1968}
{Mayer} P.,  {O'dell} C.~R.,  1968, ApJ, 153, 951

\bibitem[\protect\citeauthoryear{{Morton}}{{Morton}}{1991}]{Morton1991}
{Morton} D.~C.,  1991, ApJS, 77, 119

\bibitem[\protect\citeauthoryear{{Querci}, {Querci} \& {Kunde}}{{Querci}
  et~al.}{1971}]{Querci1971}
{Querci} F.,  {Querci} M.,    {Kunde} V.~G.,  1971, A\&A, 15, 256

\bibitem[\protect\citeauthoryear{{Riegel} \& {Crutcher}}{{Riegel} \&
  {Crutcher}}{1972}]{Riegel1972}
{Riegel} K.~W.,  {Crutcher} R.~M.,  1972, A\&A, 18, 55



\bibitem[\protect\citeauthoryear{{Snow} Jr.}{{Snow}}{1978}]{Snow1978}
{Snow} Jr. T.~P.,  1978, ApJ, 220, L93

\bibitem[\protect\citeauthoryear{{Sonnentrucker}, {Welty}, {Thorburn} \&
  {York}}{{Sonnentrucker} et~al.}{2007}]{Sonnentrucker2007}
{Sonnentrucker} P.,  {Welty} D.~E.,  {Thorburn} J.~A.,    {York} D.~G.,  2007,
  ApJS, 168, 58

\bibitem[\protect\citeauthoryear{{Souza} \& {Lutz}}{{Souza} \&
  {Lutz}}{1977}]{Souza1977}
{Souza} S.~P.,  {Lutz} B.~L.,  1977, ApJ, 216, L49

\bibitem[\protect\citeauthoryear{{Thorburn}, {Hobbs}, {McCall}, {Oka}, {Welty},
  {Friedman}, {Snow}, {Sonnentrucker} \& {York}}{{Thorburn}
  et~al.}{2003}]{Thorburn2003}
{Thorburn} J.~A.,  {Hobbs} L.~M.,  {McCall} B.~J.,  {Oka} T.,  {Welty} D.~E.,
  {Friedman} S.~D.,  {Snow} T.~P.,  {Sonnentrucker} P.,    {York} D.~G.,  2003,
  ApJ, 584, 339

\bibitem[\protect\citeauthoryear{{van Dishoeck}}{{van Dishoeck}}
  {1983}]{Dishoeck1983}
{van Dishoeck} E.~F.,  1983, Chem. Phys., 77, 277

\bibitem[\protect\citeauthoryear{{van Dishoeck} \& {Black}}{{van Dishoeck} \&
  {Black}}{1982}]{Dishoeck1982}
{van Dishoeck} E.~F.,  {Black} J.~H.,  1982, ApJ, 258, 533

\bibitem[\protect\citeauthoryear{{van Dishoeck} \& {Black}}{{van Dishoeck} \&
  {Black}}{1986}]{Dishoeck1986}
{van Dishoeck} E.~F.,  {Black} J.~H.,  1986, ApJ, 307, 332

\bibitem[\protect\citeauthoryear{{van Dishoeck} \& {Black}}{{van Dishoeck} \&
  {Black}}{1989}]{Dishoeck1989}
{van Dishoeck} E.~F.,  {Black} J.~H.,  1989, ApJ, 340, 273

\bibitem[\protect\citeauthoryear{{van Dishoeck} \& {de Zeeuw}}{{van Dishoeck}
  \& {de Zeeuw}}{1984}]{Dishoeck1984}
{van Dishoeck} E.~F.,  {de Zeeuw} T.,  1984, MNRAS, 206, 383

\bibitem[\protect\citeauthoryear{{Welty} \& {Hobbs}}{{Welty} \&
  {Hobbs}}{2001}]{Welty2001}
{Welty} D.~E.,  {Hobbs} L.~M.,  2001, ApJS, 133, 345

\end{thebibliography}

\appendix
\section{}

In this section detailed comparison between our results and those of
previously published is demonstrated. $C_2$ in interstellar clouds towards
HD\,147889, HD\,148184, HD\,169454 and HD\,179406 has already been observed,
but towards HD\,76341 and HD\,163800 our determinations are the first ones.
Comparison between our equivalent widths and data from literature is
presented in Tables \mbox{A2-A5}. There are some differences between our
results and those of previous papers which could be caused by different quality of the analysed spectra.
We used the spectra from archival database collected
by UVES, which are characterised by high resolving power \mbox{($\lambda /
\Delta \lambda \sim 85,000$)} and high signal-to-noise ratio \mbox{($S/N
\sim 200$).} The differences with published data may resemble the difficulty
of measuring of weak lines.

In Table A1 there is a comparison between our results and those from previous papers.

\begin{table*}
\begin{minipage}{150mm}
\centering
\caption{The comparison between our results and those from previous papers.
Total column densities are rescaled to the value of $f_{20}$ used in this paper.
}
\label{Table A1.}
\begin{tabular}{@{}lllllcc}
\hline
object    & $T_{kin}^a [K]$ & $T_{02} [K]$ & $T_{04} [K]$ & $T_{06} [K]$ & $N_{col}
[10^{13}\rmn{cm}^{-2}]$ &source\\
\hline
HD\,147889& $70$        &            &	         &         &$8.3\pm2.1$ &van Dishoeck \& de Zeeuw 1984\\
          &             & $52\pm22$  & $116\pm28$&         &            &Sonnentrucker et al. 2007$^b$\\
          & $39\pm2$    & $49\pm 7$  & $ 62\pm 3$&$71\pm 2$&$13.3\pm0.1$ & this work\\
HD\,148184& $40$        &            & $65$	 &         &$3.0\pm0.4$ & van Dishoeck \& de Zeeuw 1984\\
          &             &            &	         &         &$2.4$       &Danks \& Lambert 1983\\
          & $50\pm15$   & $39\pm16$  & $57\pm12$ &         &$2.8\pm0.3$ & Sonnentrucker et al. 2007$^b$\\
          & $34\pm12$   & $82\pm82$  & $65\pm12$ &$74\pm7$ & $3.0\pm0.3$ & this work\\
HD\,169454& $15$        &            &	         &         & $5.8$       & Gredel \& Muench 1986\\
          & $15^+10\,-5$&            &	         &         & $4.9\pm1.0$ & van Dishoeck \& Black 1989\\
          & $25$        &            &	         &         & $9.7$       & Gredel 1999\\
          & $20\pm5$    & $24\pm6$   & $37\pm5$  &         &             & Sonnentrucker et al. 2007$^c$\\
          & $19\pm2$     & $23\pm3$   & $31\pm1$  & $36\pm1$& $6.5\pm0.1$ & this work\\
HD\,179406& $55$        &	     &           & $66^d$  & $3.7$       & Federman \& al. 1994\\
          & 	        &            & $57\pm11$ &         &$5.0\pm0.6^f$& Sonnentrucker et al. 2007$^e$\\
          & $38\pm9$    & $38\pm 12$ & $59\pm9 $ & $62\pm5$& $3.9\pm0.1$ & this work \\
\hline
\multicolumn{7}{l}{$^a$ gas kinetic temperature $T_{kin}$ is obtained form the best-fitting model.}\\
\multicolumn{7}{l}{$^b$ calculated by Sonnentrucker et al (2007) using the equivalent widths of Dishoeck \& de Zeeuw(1984).}\\
\multicolumn{7}{l}{$^c$ calculated by Sonnentrucker et al (2007) using the equivalent widths of Gredel (1999).}\\
\multicolumn{7}{l}{$^d$ excitation temperature from all observed levels $T_{08}$}\\
\multicolumn{7}{l}{$^e$ calculated by Sonnentrucker et al (2007) using the equivalent widths of Federman et al. (1994).}\\
\multicolumn{7}{l}{$^f$ based on equivalent widths of Federman et al. 1994}
\end{tabular}
\end{minipage}
\end{table*}
The spectrum of HD\,76341 was the most noisy in our sample, and thus the
measurements are less reliable. We were able to assign 19 absorption
features to the P, Q, R lines of the $C_2$ Phillips bands in total. There we
found the lowest column density ($N_{col}=(1.1\pm0.1)\,\times 10^{13}
\rmn{cm}^{-2}$) of $C_2$ in the whole sample.

In HD\,147889 we measured the largest number of $C_2$ lines. Figure 1 shows
portions of the spectrum HD\,147889 (normalised to continuum, with removed
telluric lines) with the assignment of the interstellar $C_2$ absorption
lines. In total we were able to assign 68 absorption features to the P, Q, R
lines of the $C_2$ Phillips bands. The computed column density $(13.3\pm0.1)
\times 10^{13} \rmn{cm}^{-2}$ is the highest value of the sample.

In the case of HD\,148184 we measured 40 absorption features belonging to
the P, Q, R lines of the $C_2$ Phillips bands.

Toward HD\,163800, in total we were able to assign 29 absorption features to
the P, Q, R lines of the $C_2$ Phillips bands. We estimated
$N_{col}=(2.8\pm0.1) \times 10^{13} \rmn{cm}^{-2}$. 

For HD\,169494 we were able to identify and measure 50 absorption lines of
the P, Q, R branches of the $C_2$ Phillips bands. We estimated
$N_{col}=(6.5\pm0.1) \times 10^{13} \rmn{cm}^{-2}$ and $n_c$ about 
$330\,\rmn{cm}^{-3}$
(it is the highest value in our sample).

For HD\,179406 we measured 36 absorption features to the P, Q, R lines of
the $C_2$ Phillips bands.

\begin{table*}
\begin{minipage}{70mm}
\centering
\caption{The comparison of our equivalent widths [m\AA] to data of van Dishoeck \& de Zeeuw (1984) for HD\,147889 of the
(2,0) band.}
\label{Table A2.}
\begin{tabular}{lcc}
\hline
$B(N'')$& our results & D\&Z$^a$ \\
\hline  
 $R(6)$ & $6.1\pm0.3$& $4.0\pm0.5$\\
 $R(8)$ & $3.6\pm0.3$& $\leq 2.5$ \\
 $R(4)$ & $9.2\pm0.3$& $8.1\pm0.4$\\
 $R(10)$& $2.2\pm0.3$&            \\
 $R(2)$ &$10.5\pm0.3$& $5.3\pm0.4$\\
 $R(12)$& $1.8\pm0.3$&            \\
 $R(0)$ & $7.5\pm0.3$& $4.6\pm0.2$\\
 $Q(2)$ &$12.7\pm0.3$&$10.1\pm0.6$\\
 $Q(4)$ &$13.1\pm0.3$&$13.9\pm1.2$\\
 $P(2)$ & $2.4\pm0.3$& $2.9\pm0.3$\\
 $Q(6)$ & $9.2\pm0.3$& $7.8\pm0.6$\\
 $Q(8)$ & $6.3\pm0.3$& $4.9\pm0.7$\\
 $P(4)$ & $4.7\pm0.3$& $4.3\pm0.3$\\
 $Q(10)$& $4.5\pm0.3$& $\leq 2.5$ \\
 $P(6)$ & $5.1\pm0.3$& $3.4\pm0.3$\\
 $Q(12)$& $2.3\pm0.3$&            \\
 $P(8)$ & $2.4\pm0.4$& $\leq 2.5$ \\
 $Q(14)$& $2.1\pm0.3$&            \\
 $P(10)$& $2.2\pm0.3$&            \\
 $Q(16)$& $2.2\pm0.3$&            \\
 $P(12)$& $0.8\pm0.4$&	    \\
 $Q(18)$& $1.8\pm0.5$u&	    \\
 $Q(20)$& $2.2\pm0.7$u&	    \\
 $Q(22)$& $0.7\pm0.4$u&	    \\
 $Q(26)$& $0.7\pm0.5$u&	    \\
\hline
\multicolumn{3}{l}{\scriptsize $^a$ Measurements were made with coud{\'e} echelle spec-}\\
\multicolumn{3}{l}{\scriptsize trograph (with Reticon) which was fed by the 1.4-m}\\
\multicolumn{3}{l}{\scriptsize coud{\'e} auxiliary telescope at ESO, La Silla, Chile.}\\
\multicolumn{3}{l}{\scriptsize The resolving power in those observations was 80,000.}\\
\multicolumn{3}{l}{\scriptsize UVES allows to detect more transitions due to higher}\\
\multicolumn{3}{l}{\scriptsize resolution and S/N ratio.}
\end{tabular}
\end{minipage}
\begin{minipage}{70mm}
\caption{
The comparison of our equivalent widths [m\AA] to data of van Dishoeck \& de
Zeeuw (1984) (D\&Z) and Danks \& Lambert (1983) (D\&L) for HD\,148184 of the
(2,0) band.}
\label{Table A3.}
\begin{tabular}{lccc}
\hline
$B(N'')$ & our results & D\&Z$^a$ & D\&L$^a$\\
\hline
 $R(6)$ & $0.7\pm0.4$ & $1.3\pm0.2$ & 0.96\\
 $R(8)$ & $0.6\pm0.4$ &	      & 0.81\\
 $R(4)$ & $1.5\pm0.4$ &	      & 0.44\\
 $R(10)$& $0.7\pm0.4$ &	      & 0.48\\
 $R(2)$ & $2.7\pm0.4$ & $1.7\pm0.2$ & 1.33\\
 $R(12)$&  	        & $0.5\pm0.2$ &     \\
 $R(0)$ & $1.4\pm0.4$u& $1.4\pm0.1$ & 0.93\\
 $Q(2)$ & $3.0\pm0.4$ & $2.4\pm0.1$ & 2.41\\
 $R(14)$&        	&$\leq 0.5$   &     \\
 $Q(4)$ & $3.3\pm0.5$ & $2.5\pm0.2$ & 2.15\\
 $P(2)$ & $0.5\pm0.5$u&$\leq 0.5$   &     \\
 $Q(6)$ & $3.4\pm0.4$ & $2.4\pm0.2$ & 2.26\\
 $R(16)$&               &$\leq 0.5$   &     \\
 $Q(8)$ & $1.7\pm0.4$ & $2.0\pm0.3$ & 1.67\\
 $P(4)$ & $1.4\pm0.5$ & $0.8\pm0.2$ & 0.59\\
 $Q(10)$& $1.0\pm0.5$ & $0.8\pm0.2$ & 1.37\\
 $P(6)$ & $1.0\pm0.4$ & $1.0\pm0.1$ & 0.67\\
 $Q(12)$& $0.6\pm0.4$ & $0.9\pm0.2$ & 0.78\\
 $P(8)$ &        	& $1.0\pm0.1$ & 0.89\\
 $Q(14)$&       	& $0.6\pm0.1$ & 0.74\\
 $P(10)$&        	& $0.5\pm0.2$ & $\leq 0.67$\\
 $Q(16)$& $0.8\pm0.4$ &             &     \\
\hline
\multicolumn{4}{l}{\scriptsize $^a$ Measurements were made with coud{\'e} echelle spec-}\\
\multicolumn{4}{l}{\scriptsize trograph (with Reticon) which was fed by the 1.4-m}\\
\multicolumn{4}{l}{\scriptsize coud{\'e} auxiliary telescope at ESO, La Silla, Chile.}\\
\multicolumn{4}{l}{\scriptsize The resolving power in those observations was 80,000.}
\end{tabular}
\end{minipage}
\begin{minipage}{70mm}
\caption{The comparison of our equivalent widths [m\AA] to data of Federman
et al. (1994) for HD\,179406 of the
(2,0) band.}
\label{Table A4.}
\begin{tabular}{ccc}
\hline
$B(N'')$& our results & Federman$^a$ \\
\hline
 $R(6)$ & $1.0\pm0.5$ & $3.0\pm0.3$\\
 $R(8)$ & $0.4\pm0.4$ &	     \\
 $R(4)$ & $2.7\pm0.5$ & $3.2\pm0.2$\\
 $R(2)$ & $4.1\pm0.5$ & $5.0\pm0.4$\\
 $R(0)$ & $3.3\pm0.4$ & $2.0\pm0.2$\\
 $Q(2)$ & $4.7\pm0.4$ & $7.3\pm0.5$\\
 $R(14)$& $0.9\pm0.3$u&	     \\
 $Q(4)$ & $7.1\pm0.5$u&$5.7\pm0.5$ \\
 $P(2)$ & $1.4\pm0.4$ &$\leq 1.1$  \\
 $Q(6)$ & $3.6\pm0.4$ & $4.2\pm0.4$\\
 $Q(8)$ & $1.3\pm0.8$u&$1.1\pm0.3^b$\\
 $P(4)$ & $0.9\pm0.5$ &$1.8\pm0.3^b$\\
 $Q(10)$& $2.2\pm0.5$u&	     \\
 $P(6)$ & $1.7\pm0.5$u&$3.1\pm0.6^c$\\
 $Q(14)$& $0.6\pm0.3$u&	     \\
\hline
\multicolumn{3}{l}{\scriptsize $^a$ Measurements were made with coud{\'e} spectrometer}\\
\multicolumn{3}{l}{\scriptsize which was on the 2.1-m telescope at McDonald}\\
\multicolumn{3}{l}{\scriptsize Observatory. The resolving power in those}\\
\multicolumn{3}{l}{\scriptsize observations typically was 25,000 to 30,000.}\\
\multicolumn{3}{l}{\scriptsize $^b$ $P(4)$ and $Q(8)$ lines are blended (Federman
et al. 1994).}\\
\multicolumn{3}{l}{\scriptsize $^c$ Result is suspect because of possible stellar}\\
\multicolumn{3}{l}{\scriptsize contamination (Federman et al. 1994).}
\end{tabular} 
\end{minipage}
\begin{minipage}{70mm}
\caption{
The comparison of our equivalent widths [m\AA] to data of Gredel \& Muench
(1986) (G\&M), van Dishoeck \& Black (1989) (D\&B) Crawford (1997) (Cr) and
Gredel (1999) (G) for HD\,169454 of the (2,0) band.}
\label{Table A5.}
\begin{tabular}{lcclcc}
\hline
$B(N'')$& our results& G\&M$^a$ & D\&B$^a$ & Cr$^b$ & G$^c$\\
\hline  
 $R(6)$ &$2.4\pm0.3$ &           & & & $\leq 2$\\
 $R(8)$ &$0.7\pm0.3$ &           & & & $\leq 1$\\
 $R(4)$ &$3.7\pm0.3$ &$3.8\pm0.8$& & & $4.1\pm1$\\
 $R(10)$&$0.4\pm0.3$u& 	   & & & \\
 $R(2)$ &$8.1\pm0.3$ &$6.1\pm1.2$&$6.0\pm0.5$& & $8.6\pm1$\\
 $R(12)$&$0.3\pm0.2$ & 	   & & &\\
 $R(0)$ &$8.5\pm0.3$ &$8.2\pm0.8$&$5.6\pm1.0$&$6.23\pm0.50$ & $8.2\pm1$\\
 $Q(2)$ &$10.3\pm0.3$&$8.0\pm1.6$&$6.6\pm0.8$&$8.14\pm1.18$ & $10.3\pm1$\\
 $Q(4)$ &$7.9\pm0.3$ &$5.0\pm1.0$&$7.9\pm0.7$&$3.98\pm1.03$ & $9.7\pm1$\\
 $P(2)$ &$2.0\pm0.3$ &$2.3\pm0.5$&$1.3\pm0.3$&$2.82\pm0.55$ & $2.7\pm2$\\
 $Q(6)$ &$3.0\pm0.3$ &$3.1\pm0.6$&$3.9\pm0.5$&$2.89\pm0.59$ & $3.8\pm1$\\
 $Q(8)$ &$2.3\pm0.3$ &$2.6\pm1.0$&$1.5\pm0.5$&$1.28\pm0.59$ & $3.2\pm1.5$\\
 $P(4)$ &$1.7\pm0.3$ &$2.0\pm0.8$&$1.7\pm0.8$&$1.68\pm0.62$ & $3.4\pm1.5$\\
 $Q(10)$&$0.8\pm0.3$ &$1.0\pm0.5$&$1.0\pm0.8$& & $\leq 1$\\
 $P(6)$ &$0.9\pm0.8$ &$1.9\pm0.8$&$< 1.5$& & $3.0\pm1$\\
 $Q(12)$&$1.0\pm0.3$ &$1.0\pm0.5$&$< 1.5$& & $\leq 1$\\
 $P(8)$ &$0.8\pm0.3$ &      & & & \\
 $Q(14)$&$0.6\pm0.4$ &      & & &\\
 $Q(16)$&$0.6\pm0.4$ &      & & &\\
\hline
\multicolumn{6}{l}{\scriptsize $^a$ Measurements were made with coud{\'e} echelle spectrograph (with Reticon)}\\
\multicolumn{6}{l}{\scriptsize which was fed by the 1.4-m coud{\'e} auxiliary telescope at ESO, La Silla,}\\
\multicolumn{6}{l}{\scriptsize Chile. The resolving power in those observations was 80,000.}\\
\multicolumn{6}{l}{\scriptsize $^b$ The observations were obtained with the Ultra-High-Resolution Facility}\\
\multicolumn{6}{l}{\scriptsize (R = 910,000) at the Anglo-Australian Telescope}\\
\multicolumn{6}{l}{\scriptsize $^c$  The observations were made using the ESO New Technology Telescope,}\\
\multicolumn{6}{l}{\scriptsize La Silla, Chile, R$\sim$45,000.}\\
\end{tabular}
\end{minipage}
\end{table*}

\label{lastpage}
\end{document}